\documentclass[12pt]{amsart}
\usepackage[dvips]{color}
\usepackage{amsmath}
\usepackage{amsxtra}
\usepackage{amscd}
\usepackage{amsthm}
\usepackage{amsfonts}
\usepackage{amssymb}
\usepackage{eucal}
\usepackage{epsfig}
\usepackage{graphics}
\def\({\left(}
\def\){\right)}

\newcommand{\mub}{\mbox{\boldmath$\mu$}}
\newcommand{\vphis}{\scalebox{.7}{\boldmath$\varphi$}}
\newcommand{\vphi}{\mbox{\boldmath$\varphi$}}
\newcommand{\betab}{\mbox{\boldmath$\beta$}}
\newcommand{\gammab}{\mbox{\boldmath$\gamma$}}
\newcommand{\cb}{\mathbf{c}}
\newcommand{\bb}{\mathbf{b}}

\newcommand{\tb}{\mathbf{t}}

\newcommand{\nn}{\nonumber}
\newcommand{\bea}{\begin{eqnarray}}
\newcommand{\ena}{\end{eqnarray}}
\def\bel{\begin{eqnarray}}
\def\enl{\end{eqnarray}}
\newcommand{\be}{\begin{eqnarray*}}
\newcommand{\en}{\end{eqnarray*}}
\newcommand{\ba}{\begin{array}}
\newcommand{\ea}{\end{array}}

\newcommand{\Tr}{{\rm Tr}}

\newenvironment{tenumerate}{
  \begin{enumerate}
  
  }{\end{enumerate}}
\newcommand{\bi}{\begin{tenumerate}}
\newcommand{\ei}{\end{tenumerate}}
\newcommand{\isoto}[1][]%
{{\mathop{\buildrel{\sim}\over\longrightarrow}\limits_{#1}}}

\def\[{\left[}
\def\]{\right]}
\newcommand{\la}{\lambda}

\newcommand{\al}{\alpha}

\newcommand{\z}{\zeta}

\newcommand{\om}{{\omega}}

\numberwithin{equation}{section}

\newcommand{\ob}{\mbox{\boldmath$\omega $}}

\def\half{\textstyle{\frac  1 2}}

\def\bi{\mathbf{i}}

\def\Io{I_\mathrm{odd}(m)}
\newcommand{\zbz}{z,\bar{z}}

\begin{document}
\begin{title}{Hidden Grassmann structure in the XXZ model V: 
sine-Gordon model.}
\end{title}
\author{M.~Jimbo, T.~Miwa and  F.~Smirnov}
\address{MJ: Department of Mathematics, 
Rikkyo University, Toshima-ku, Tokyo 171-8501, Japan}
\email{jimbomm@rikkyo.ac.jp}
\address{TM: Department of 
Mathematics, Graduate School of Science,
Kyoto University, Kyoto 606-8502, 
Japan}\email{tmiwa@math.kyoto-u.ac.jp}
\address{%FS\footnote{Walton Professor}${}^,$
FS\footnote
{Membre du CNRS}: 
Hamilton Mathematical Institute and School of Mathematics,
Trinity College, Dublin 2, Ireland
\newline
Laboratoire de Physique Th{\'e}orique et
Hautes Energies, Universit{\'e} Pierre et Marie Curie,
Tour 13, 4$^{\rm er}$ {\'e}tage, 4 Place Jussieu
75252 Paris Cedex 05, France}\email{smirnov@lpthe.jussieu.fr}

\begin{abstract}

We study one-point functions of the sine-Gordon model on a cylinder.  
Our approach is based on a fermionic description of the space of descendent fields, 
developed in our previous works for conformal field theory 
and the sine-Gordon model on the plane. 
In the present paper we make an essential addition 
by giving a connection between various primary fields in terms of yet another 
kind of fermions. 
The one-point functions of primary fields and descendants 
are expressed in terms of a single function
$\omega_R^\mathrm{sG}(\z,\xi|\alpha)$, 
defined via the data from the thermodynamic Bethe Ansatz
equations. \end{abstract}

\maketitle
%\vskip.2cm
%\noindent{\bf Mathematics Subject Classification (2010). } 81T40.
%\vskip .2cm
%\noindent {\bf Keywords.} quantum field theory, integrable models, sine-Gordon model, conformal field theory.
\section{Introduction}\label{intro}

The one-point functions are important data in Quantum Field Theory (QFT).
Indeed, if one applies the Operator Product Expansion (OPE) 
for computing ultra-violet asymptotics of correlation functions, 
two objects are
needed: the coefficients of the OPE and the one-point functions. 
The coefficients of the OPE are purely ultra-violet data which
have nothing to do with the infra-red environment of the system.
In principle, these coefficients are governed by the
convergent perturbation theory based on the ultra-violet Conformal
Field Theory (CFT). For the one-point functions the situation is
different. These are the data which depend essentially
on the
infra-red environment, and cannot be obtained from the 
ultra-violet CFT. So, in order to find them
one has to develop methods different from CFT.

Let us illustrate these general ideas on a particular example of two-dimensional
Integrable Field Theory (IFT) 
---
the famous sine-Gordon (sG) model. In this paper we consider the sG model not as a
perturbation of $c=1$ CFT, but rather as a perturbation of 
the complex Liouville model. We emphasise this point of view
writing the Euclidean action as
\begin{align}
\mathcal{A}^\mathrm{sG}&=\int \left\{ \Bigl[\frac 1 {4  \pi} \partial _z\vphi (z,\bar{z})\partial_{\bar{z}}\vphi (z,\bar{z})-
\frac{\mub ^2}{\sin\pi\beta ^2} e^{-i\beta\vphis(z,\bar{z})}\Bigr]\right.
\label{action1}\\ &\quad\quad\quad\quad\quad\quad\quad\quad\quad\quad\quad\quad\quad \left.
-
\frac{\mub ^2}{\sin\pi\beta ^2}   e^{i\beta\vphis(z,\bar{z})}    \right\}\frac{idz\wedge d\bar{z}}2\,,
\nn
\end{align}
where the normalisation of the dimensional coupling
constant $\mub^2$ is 
chosen
for future convenience. 
For historical reasons we shall mostly use the parameter
\begin{align}
\nu =1-\beta ^2
\label{nu}
\end{align}
as the coupling constant, in term of which
the semiclassical domain is $\nu \simeq 1$. 
The complex Liouville model is conventionally identified with the
minimal model of CFT, and the perturbing field 
$e^{i\beta\vphis(\zbz)}$ is called $\Phi _{1,3}
(\zbz)$. The central charge
is
$$c=1-6\frac {\nu ^2}{1-\nu}\,.$$

We shall consider the 
Euclidean correlation functions on 
an 
infinite cylinder
of circumference $2\pi R$. 
We write the point $(z,\bar{z})=(0,0)$ simply as $0$.
The conformal map $w=e^{-z}$ brings the cylinder to the Riemann sphere. 
We use the symbol $(z,\bar{z})=-\infty,\infty$ to represent 
the points corresponding to $w=\infty,0$ on the latter.
We use the name  {``}primary fields" for
the 
exponential fields, which are parametrised as
$$ 
\Phi _\al(\zbz)=e^{
\ \frac { \nu}{2(1-\nu)}
\al \left\{ {i\beta} \vphis (\zbz)\right\}}\,. 
$$
Its scaling dimension is $2\Delta_{\al}$ where 
\begin{align*}
\Delta_{\alpha}=\frac{\nu^2}{4(1-\nu)}\alpha(\alpha-2)\,.
\end{align*}
So, our 
normalisation is such that the shift of $\al$ by $\frac {2(1-\nu)}\nu$ corresponds in the CFT language to normal ordered multiplication
by the primary field $\Phi _{1,3}$. 
We  consider $\Phi _\al(\zbz)$ as a
primary field for the model %with 
given by the
action \eqref{action1}. 
We work with the range 
$$ 
0<\al<2\,,
$$
and when needed, treat physical objects such as correlation functions  
as analytic continuation from this domain.  
The analyticity in $\alpha$ is a requirement built 
implicitly in our definition of the model. The case $\al =0$ is special, and we shall comment on it in subsection
\ref{al=0}.

The OPE for the product of two primary fields in the sG model looks 
as follows:
\begin{align}
\Phi _{\al _1}(z,\bar{z})\Phi _{\al _2}(0)&=\sum\limits _{m=-\infty}^{\infty}
\sum\limits _{N,\bar{N}} \(\mub ^2 r^{2\nu}\)^{|m|}
C_{\al _1,\al _2}^{m, N,\bar{N}}\(\mub ^4r^{4\nu}\)\label{OPE}\\
&\times r^{\frac {\nu ^2}{1-\nu}\al _1\al_2
+2m^2(1-\nu)+2\al m\nu     }\ z^{|N|}\bar{z}^{|\bar{N}|}\ \mathbf{l}_{-N}
\bar{\mathbf{l}}_{-\bar{N}}\Phi _{\al+2m\frac{1-\nu}\nu}(0)\,,\nn
\end{align}
where $\al =\al_1+\al _2$, $r=\sqrt{z\bar{z}}$. The formula 
\eqref{OPE} needs some comments. In order
to avoid resonances  we impose the requirement of
incommensurability:
$\nu$, $\nu\alpha_i$ and $1$
are linearly independent over $\mathbb{Q}$.
By $\mathbf{l}_{-N}\bar{\mathbf{l}}_{-\bar{N}}\Phi _{\al+2\frac{1-\nu}\nu}(0)$ 
we mean the unique operator in the sG theory, which tends to the 
corresponding Virasoro descendant in the conformal limit, and does not contain
finite counterterms.
Following  \cite{HGSIV}, we use the letters 
$\mathbf{l}_n$ and $\bar{\mathbf{l}}_n$
to denote the Virasoro generators  acting on fields 
at $(z,\bar{z})=0$, to make distinction %with 
from 
the Fourier components of the CFT energy-momentum tensor acting on the states
at $(z,\bar{z})=\pm\infty$.
For 
$N=\{n_1,\cdots ,n_p\}$, $\bar{N}=\{\bar{n}_1,\cdots ,\bar{n}_q\}$ 
with $n_1\geq\cdots\geq n_p>0$, $\bar n_1\geq\cdots\geq\bar n_q>0$, we set
$$
\mathbf{l}_{-N}=\mathbf{l}_{-n_1}\cdots\mathbf{l}_{-n_p},
\quad 
\bar{\mathbf{l}}_{-\bar{N}}= \bar{\mathbf{l}}_{-\bar{n}_1}\cdots
 \bar{\mathbf{l}}_{-\bar{n}_q},
\quad |N|=\sum_{k=1}^p{n_k},\quad  |\bar{N}|=\sum_{k=1}^q{\bar{n}_k}\,.
$$

 The structure
functions $C_{\al _1,\al _2}^{m, N,\bar{N}}(t)$ are 
 power series in $t$. It is assumed \cite{FFLZZ}
 that the series converge. But independently on this assumption
 we shall consider $C_{\al _1,\al _2}^{m, N,\bar{N}}(t)$ as something
 known because the coefficients of the series can be expressed
 through the Coulomb gas integrals. 

 We want to use the OPE \eqref{OPE} for the calculation of the 
 two-point functions on the cylinder. The main problem is to compute
 the one-point functions of 
 $\mathbf{l}_{-N}
\bar{\mathbf{l}}_{-\bar{N}}\Phi _{\al+2\frac{1-\nu}\nu}(0)$. 
We follow
the idea of \cite{OP} in order to solve this problem.
Namely,  we use
the fermionic basis introduced for CFT
in \cite{HGSIV} following
the study of lattice models \cite{HGS,HGSII,HGSIII}. Let us explain
this point. 

The basic idea of \cite{HGS,HGSII,HGSIII,HGSIV} is 
to consider operators acting on the space of operators, 
rather than those acting on the space of states. 
The Virasoro algebra serves as a tool for labeling local fields
in the perturbed theory, even though the conformal invariance is broken.
So the space of descendants of the exponential field  $\Phi _\al (0)$ in the sG model  
is identified as a linear space 
with the tensor product of Verma modules $\mathcal{V}_\al\otimes
\overline{\mathcal{V}}_\al$. 
With this identification,  
the operators acting on the sG local fields can be represented as ones
acting on Verma modules. 
Let us give a simplest example.

Regard the compact direction on the cylinder as time. 
It is well known that the sG model possesses infinitely
many local integrals of motion $I_{2k-1}$, $\bar{I}_{2k-1}$, 
which include in particular
the Hamiltonian $H=I_1+\bar{I}_1$ and the momentum 
$P=I_1-\bar{I}_1$. 
They act on local operators by commutators. 
We denote this action by $\mathbf{i}_{2k-1}$, $\bar{\mathbf{i}}_{2k-1}$. 
In our euclidean approach, the commutator is represented as 
the difference of local densities integrated over 
$\mathbb{R}+i0$ and $\mathbb{R}-i0$.
It is intuitively clear that
the one-point functions of the field obtained by this action always vanish: 
by moving the contours along the compact direction the integrals cancel with each other.  
%In order that it is possible 
To make it work 
the boundary conditions at $z=\pm\infty$ are chosen appropriately 
(see Section 2 for the relevant discussion). 
From the very definition of Zamolodchikov's construction \cite{zam}, 
%upon the identification of the sG descendants with 
the operators $\mathbf{i}_{2k-1}$, $\bar{\mathbf{i}}_{2k-1}$ act 
on $\mathcal{V}_\al\otimes\overline{\mathcal{V}}_\al$ as 
$$\mathbf{i}_1=\mathbf{l}_{-1},\quad 
\mathbf{i}_3=\sum\limits _{n=0}^\infty \mathbf{l}_{-n-2}
\mathbf{l}_{n-1},\quad etc. \,,$$
and similarly for the second chirality. 
For our goal, however, this nice construction is useless 
since the one-point functions of these descendants vanish.   
%We have to factor out the action of the local
%integrals of motion and consider
It means that the one-point function is a 
linear functional defined on the quotient space
$$
\mathcal{V}^\mathrm{quo}_\al=
\mathcal{V}_\al\ /\ \sum_{k=1}^\infty\mathbf{i}_{2k-1}\mathcal{V}_\al,
\quad \overline{\mathcal{V}}^\mathrm{quo}_\al=
\overline{\mathcal{V}}_\al\ /\ \sum_{k=1}^\infty
\bar{\mathbf{i}}_{2k-1}\overline{\mathcal{V}}_\al\,.
$$
Notice that logically this procedure is the same as considering the 
cohomologies of the affine Jacobi variety in the classical
case \cite{NakaSm,BBS}. 

Now we proceed to a construction more interesting to us. 
In the paper \cite{HGSIII} it was shown, 
for the six-vertex model on the cylinder, that 
the expectation values of quasi-local operators can be simplified
by considering certain fermions acting on them. 
As explained in \cite{OP}, this 
construction can be easily generalised to an inhomogeneous six-vertex model. 
Considering the CFT and the sG models as scaling limits of homogeneous and inhomogeneous six-vertex models, respectively,
we conclude that the action of our fermions
allows the same identification as the action of the local
integrals of motion.
Let us be more specific. 
We have two kinds of fermions $\betab ^*_{2j-1}$ and $\gammab^*_{2j-1}$.
The indices agree with the scaling dimension in CFT.
The bilinear combinations $\betab ^*_{2j-1}\gammab ^*_{2j-1}$
act from $\mathcal{V}_\al$ to itself, and similarly for the second chirality. 
%Consider the descendants of the
%primary field $\Phi _{\al}(0)$. It is explained in \cite{OP} that
Altogether the quotient space $\mathcal{V}^\mathrm{quo}_\al\otimes
\overline{\mathcal{V}}^\mathrm{quo}_\al$
allows a fermionic basis:
\begin{align}
\betab^* _{I^+}\bar{\betab} ^*_{\bar{I}^+}
\bar{\gammab }^*_{\bar{I}^-}\gammab^*_{I^-}
\Phi _{\al}(0)\,,\label{basis}
\end{align}
where $\#(I^+)=\#(I^-)$, $\#(\bar I^+)=\#(\bar I^-)$. We set generally
\begin{align}
&I^+=\{2i^+_1-1,\cdots , 2i^+_p-1\}\ (1\leq i^+_1<\cdots<i^+_p),\quad
\betab^* _{I^+}=\betab^*_{2i^+_1-1}\cdots \betab^*_{2i^+_p-1}\,,
\label{mult1}\\
&I^-=\{2i^-_1-1,\cdots , 2i^-_q-1\}\ (1\leq i^-_1<\cdots<i^-_q),\quad
\gammab^* _{I^-}=\gammab^*_{2i^-_q-1}\cdots \gammab^*_{2i^-_1-1}\,,
\nn
\end{align}
and similarly for the second  chirality. %Recall that
%$\betab^*_{2i-1}$, $\gammab^*_{2i-1}$, $\bar{\betab}^*_{2i-1}$,
%$\bar{\gammab}^*_{2i-1}$ are fermions. 

The quotient space 
$\mathcal{V}^\mathrm{quo}_\al\otimes
\overline{\mathcal{V}}^\mathrm{quo}_\al$ 
can also be realised as a module
created from $\Phi _\al(0)$ by the action of
$\mathbf{l}_{-2k}$, $\bar{\mathbf{l}}_{-2k}$.
The fermionic basis \eqref{basis}
can be related to this basis 
of $\mathcal{V}^\mathrm{quo}_\al\otimes
\overline{\mathcal{V}}^\mathrm{quo}_\al$  as is explained in \cite{HGSIV}. 
On the other hand, for the operators 
\eqref{basis} the one-point functions can be computed. 
This has been done in \cite{OP} for the sG model on the plane. 
We shall show that the case of a cylinder can be dealt with
by a simple generalisation.

However, there is an important difference from the paper \cite{OP}.
There we did not need to compute the one-point functions of the primary fields
because they have been known \cite{LukZam}. Now the situation
is different. These one-point functions are unknown
on the cylinder, but they are needed for the application to OPE.
The main new result of the present paper is that we
found a fermionic description for them.

First, let us correct one mistake done in the paper \cite{HGSIV}.
There we seriously considered only the quadratic expressions 
$\betab ^*_{2j-1}\gammab ^*_{2k-1}$.  For the individual operator $\betab^*_{2j-1}$
it was erroneously stated that it
acts from $\mathcal{V}_\al $ to $\mathcal{V}_{\al +2\frac{1-\nu}{\nu}}$. 
Obviously, this is impossible for dimensional reasons. 
As usual in CFT one has to introduce the screening
operators in order to correct this. 
Surprisingly enough, these
screening operators can be constructed from the same
material as $\gammab^*_{2j-1}$. 
Let us be more precise.
The operators $\gammab^*_{2j-1}$ are obtained as coefficients
of the asymptotical expansion at $\la =\infty$ of a generating
function $\gammab^*(\la)$, where the asymptotics goes in the
powers $\la ^{-\frac {2j-1}{\nu}}$. 
In the weak sense the
function  $\gammab^*(\la)$ is analytical, 
and we can define its
expansion at $\la =0$ which goes in powers $\la ^{-\al+2j}$.
The corresponding coefficients are denoted $\gammab ^*_{\mathrm{screen}, j}$. 
Similarly the operators $\bar{\betab^*}_{2j-1}$
are the coefficients of the asymptotics of  $\bar{\betab}^*(\la)$ at $\la=0$,  
and we introduce $\bar{\betab} ^*_{\mathrm{screen}, j}$ as coefficients
of its power series at $\infty$ (the series goes in $\la ^{\al-2j}$).

%Namely, using the
%scaling limit of the same fermionic operators of the lattice theory
%we define new operators
%$\gammab ^*_{\mathrm{screen}, j}$ and $\bar{\betab}^*_{\mathrm{screen},j}$ for
%$j=1,2,\cdots$. Using them
%we create primary fields
%so that their one-point functions become calculable.

Let us introduce some notation. 
We use the multi-index 
$$I(m)=\{1,2,\cdots ,m\}\,,$$
and define 
\begin{align}\gammab ^*_{\mathrm{screen}, I(m)}=
\gammab ^*_{\mathrm{screen}, m}\cdots\gammab ^*_{\mathrm{screen}, 1}\,,\quad
\bar{\betab}^*_{\mathrm{screen},I(m)}=
\bar{\betab}^*_{\mathrm{screen},1}\cdots \bar{\betab}^*_{\mathrm{screen},m}\,.\label{mult2}\end{align}
Then the 
$m$-fold
screened primary field is by
definition
\begin{align}
\Phi _{\al}^{(m)}(0)=
i^{m}\mub^{2m}\prod\limits _{j=1}^{m}\cot {\textstyle \frac{\pi\nu}2(2j-\al)
\times
\bar{\betab}^*_{\mathrm{screen},I(m)}\ \gammab ^*_{\mathrm{screen}, I(m)}%\cdots \bar{\betab}^*_{\mathrm{screen},m}
\Phi _{\al}(0)\,,}
\label{phim}
\end{align}
where the multiplier in the right hand side is introduced for future
convenience. 

We claim that the basis of the quotient space
$\mathcal{V}^\mathrm{quo}_{\al+2m\frac {1-\nu}\nu}\otimes
\overline{\mathcal{V}}^\mathrm{quo}_
{\al+2m\frac {1-\nu}\nu}$  for $m > 0$ can be
constructed as
\begin{align}
\betab^* _{I^+}\bar{\betab} ^*_{\bar{I}^+}
\bar{\gammab }^*_{\bar{I}^-}\gammab^*_{I^-}
\Phi _{\al}^{(m)}(0)\,,\label{sss}
\end{align}
where $\#(I^+)=\#(I^-)+m$, $\#(\bar I^-)=\#(\bar I^+)+m$.

This is not an abstract statement, but
the identification is done by explicit formulae.
In particular, taking the element of the lowest dimension
we obtain the primary field:
\begin{align}
\Phi _{\al +2m\frac {1-\nu}\nu}(0)\cong C_m(\al)
\betab^*_{\Io}
\bar{\gammab}^*_{\Io}
\Phi ^{(m)}_{\al}(0)\,,
\label{shift-primary}
\end{align}
where $C_m(\al)$ is an important constant (see \eqref{calpha} below), 
and
$$\Io=2I(m)-1=\{1,3,\cdots , 2m-1\}\,.$$  
From now on we use the following notation
for multi-indices
\begin{align}
\mathrm{if} \quad J=\{j_1,\cdots ,j_p\}\quad
\mathrm{then}\quad aJ+b=\{aj_1+b,\cdots ,aj_p+b\}\,.\label{convmi}
\end{align} 
The operators on the two sides of \eqref{shift-primary} belong 
 {\it a priori} to different spaces.
 By the symbol  $\cong$ we imply that 
the two vectors from different spaces act as the same local operator in CFT.
This statement 
is not a mathematical theorem, because we cannot check completely
the identification.  What we can compute are the three-point functions
with two primary fields of the same scaling dimension
$\Delta _{\kappa+1}$. The main result of our fermionic construction
is that for the right 
hand side of 
\eqref{shift-primary} this three-point function
can be evaluated in terms of one 
function $\omega ^\textrm{sc}(\la,\mu|\kappa,\kappa,\al)$. We computed
the asymptotics of this function for $\kappa\to\infty$ up to $\kappa^{-8}$.
In the left hand side we use for the three-point function Dotsenko-Fateev formula \cite{DF}.
The comparison of the asymptotics goes
in an amazingly nice way. There are certain consistency conditions which we also check. 
We compute also the constant $C_m(\al)$ and explain its relation
to the Lukyanov-Zamolodchikov one-point function.

Consider the part of the sum in the right hand side of \eqref{OPE} corresponding to descendants
of the primary field $\Phi _{\al +2m \frac {1-\nu}\nu}(0)$, $m\ge 0$. Modulo the descendants generated
by the local integrals of motion it can be rewritten in the fermionic basis:
\begin{align}
&\(\mub ^2 r^{2\nu}\)^{m}r^{
2m^2(1-\nu)+2\al m\nu     }
\sum\limits _{N,\bar{N}}C_{\al _1,\al _2}^{m, N,\bar{N}}\(\mub ^4r^{4\nu}\)
\ z^{|N|}\bar{z}^{|\bar{N}|}\ \mathbf{l}_{-N}
\bar{\mathbf{I}}_{-\bar{N}}\Phi _{\al+2m\frac{1-\nu}\nu}(0)\label{OPEferm}\\
&\cong
\(\mub ^2 r^{2\nu}\)^{m}r^{
-2\nu m^2+2\al m\nu     }\sum\limits _{{I,J,\bar{I},\bar{J}}\atop 
{\#(I^+)=\#(I^-)+m,\ \#(\bar{I}^-)=\#(\bar{I}^+)+m}}\widetilde{C}_{\al _1,\al _2}^{I^+,I^-,\bar{I}^+,\bar{I}^-}
\(\mub ^4r^{4\nu}\)\nn\\
&\times z^{|I^+|+|I^-|}\bar{z}^{|\bar{I}^+|+|\bar{I}^-|}
\betab^*_{I^+}\ \bar{\betab}^*_{\bar{I}^+}
\ \bar{\gammab}^*_{\bar{I}^-}\ \gammab^*_{I^-}
\ \Phi ^{(m)}_{\al}(0)\,.\nn
\end{align}
The 
structure
functions $\widetilde{C}_{\al _1,\al _2}^{I,J,\bar{I},\bar{J}}(t)$ can be in principle
recalculated from $C_{\al _1,\al _2}^{m, N,\bar{N}}(t)$ because the relation between the usual
basis of Verma module and the fermionic basis can be obtained
\cite{HGSIV}. 
Nevertheless,  we
believe that it should be 
possible to find a direct way to compute 
$\widetilde{C}_{\al _1,\al _2}^{I^+,I^-,\bar{I}^+,\bar{I}^-}(t)$. 
For, the elements of the fermionic basis have very simple 
one-point functions, 
and  hence the structure functions
associated with them  should also have a fundamental meaning.
Let us now turn to the discussion of the one-point functions.

The one-point functions of $\betab^* _{I^+}\bar{\betab} ^*_{\bar{I}^+}
\bar{\gammab }^*_{\bar{I}^-}\gammab^*_{I^-}
\Phi _{\al}^{(m)}(0)$ are given by determinant formulae. We introduce the function
$\omega ^{\mathrm{sG}}_R(\la,\mu|\al)$ by a TBA-like equation written in Section \ref{TBA}.
 Equation of this kind 
was written for the first time in \cite{BGKS}.

It is very convenient to use instead of $\omega ^{\mathrm{sG}}_R(\la,\mu|\al)$ the
function $\Theta ^{\mathrm{sG}}_R(l,j|\al)$ which is related to the Mellin transform of 
$\omega ^{\mathrm{sG}}_R(\la,\mu|\al)$ as in \eqref{finomega}.
Then our main formula reads
\begin{align}
&\frac {\langle
\betab^* _{I^+}\bar{\betab} ^*_{\bar{I}^+}
\bar{\gammab }^*_{\bar{I}^-}\gammab^*_{I^-}
\Phi_{\al}^{(m)}(0)
 \rangle^\mathrm{sG}_R}
{\langle\Phi_{\al }(0) \rangle^\mathrm{sG}_R}\label{themainintro}\\
&=
\mub^{2m\al -2m^2 +\frac 1 {\nu}\(|I^+|+|I^-|+|\bar{I}^+|+|\bar{I}^-|\)}\nn
\ \mathcal{D}^{\mathrm{sG}}_R\(I^+\cup(- \bar{I}^+)\ |\ I^-\cup (-\bar{I}^-)|\al\)\,,\nn
\end{align}
where for $A=\{a_j\}_{j=1,\cdots,n}$, $B=\{b_j\}_{j=1,\cdots,n}$ we set
\begin{align}
&\mathcal{D}^{\mathrm{sG}}_R(A|B|\al)=\prod\limits
_{j=1}^n \mathrm{sgn}(a_j) \mathrm{sgn}(b_j)\nn\\
&\times\(\frac i{2\pi\nu^2}\)^n\det\left.
\( \Theta ^{\mathrm{sG}}_R\(
{\textstyle \frac {ia_j}{2\nu}},{\textstyle \frac {ib_k}{2\nu}}|\al\)
-\mathrm{sgn}(a_j)\delta _{a_j,-b_k}2\pi\nu\cot
{\textstyle \frac {\pi}{2\nu }}(a_j+\nu\al)\)\right|_{j,k=1,\cdots ,n}\,.\nn
\end{align}
The mass of soliton  $M$  is related to $\mub$ by
the famous formula
\cite{alzam}:
\begin{align}
 \mub
 =\left[ M\frac {\sqrt{\pi} \ \Gamma (\frac 1 {2\nu})}
{2\Gamma (\frac {1-\nu}{2\nu})  } \Gamma (\nu)^{-\frac 1 \nu} \right]^{\nu}\,.\label{musol}
\end{align}
Notice that in this paper we do not use this formula as an input, 
but it comes as a result of our computations. 
The operators $\gammab^*_{\mathrm{screen},j}$, $\bar{\betab}^*_{\mathrm{screen},j}$ are very important in the CFT
computations. However, their contribution to the formula \eqref{themainintro} 
enters as a decoupled diagonal
block in the determinant, and 
cancels with the multiplier
in the definition of $\Phi _{\al}^{(m)}(0)$ \eqref{phim}
which is 
introduced exactly for this reason.

The formula \eqref{themainintro} requires certain consistency relations
for the function $\Theta ^{\mathrm{sG}}_R\(l,j|\al\)$
because one can obtain the primary field $\Phi _{\al +2m\frac {1-\nu}\nu}$
either directly from $\Phi _{\al }$ or in $m$ steps passing through 
$\Phi _{\al +2k\frac {1-\nu}\nu}$ $(k=1,\ldots,m-1)$. 
One can also write a formula (see \eqref{IDENDESC}) for the descendants similar to
\eqref{shift-primary} for the primary field; it requires yet another consistency relation.
We write these relations down and prove them in Section \ref{cons}.

From the computations of Section \ref{cons}  we draw one
more conclusion. Until now we considered only the descendants
of $\Phi _{\al +2m \frac{1-\nu}\nu}$ for $m\ge 0$. The question is 
what to do with $\Phi _{\al +2m \frac{1-\nu}\nu}$ for $m < 0$
which are also present in \eqref{OPE}. The direct way to tackle this
problem consists in considering the domain $0>\al>-2$ and 
analytically continuing from there. But actually this is not needed.
The consistency relations imply the existence 
of the screened
primary field $\Phi _{\al}^{(m)}$
for $m<0$  such that
$$\Phi _{\al +2m\frac {1-\nu}\nu}(0)\cong C_{m}(\al)
\betab^*_{\Io}
\bar{\gammab}^*_{\Io} \Phi ^{(m)}_{\al}(0)\,,$$
where $\Io=-I_\mathrm{odd}(-m)$, and
the constant $C_{m}(\al)$ for negative $m$ is given in Section \ref{cons} (see \eqref{Cnegative}).
A basis of the corresponding Verma module is
constructed as 
\begin{align}
\betab^* _{I^+}\bar{\betab} ^*_{\bar{I}^+}
\bar{\gammab }^*_{\bar{I}^-}\gammab^*_{I^-}
\Phi _{\al}^{(m)}(0)\,,
\end{align}
where $\#(I^+)=\#(I^-)+m$, $\#(\bar{I}^-)=\#(\bar{I}^+)+m$.
The difference from
the previous case is that the number
of $\betab^*$'s ($\bar{\gammab}^*$'s) is smaller than the number
of $\gammab^*$'s ($\bar{\betab}^*$'s). 
It is useful to keep in mind the analogy
with the Dirac fermions with different vacua. 
Then the parameter $\al$ which counts the
primary fields is nothing but the value of the zero-mode.  
This reminds us very much of the construction in the paper \cite{BBS}. 

The plan of the paper is as follows. In Section 2
we discuss the scaling limit of the XXZ model in 
the  homogeneous and 
the inhomogeneous cases which lead respectively to 
CFT
and the sG model. Section 3 deals with the fermionic construction
for chiral CFT. Here 
we introduce the screening fermions. In section 4 we present the fermionic construction of primary fields.
We explain the construction of descendants in Section 5. 
In Section 6 we compare the three-point functions
constructed through fermions with 
the usual CFT formulae.
In section 7 we continue the fermionic description of
the sG model and present  the pairings of fermions. 
The main formula for the one-point functions is given in Section 8.
Section 9 deals with the crucial issue of consistency. In Section 10
we show agreement of our results with several known formulae. 
Some concluding remarks are given in Section 10. In Appendix A 
we provide necessary formulae concerning the three-point
functions in CFT. In Appendix B  several terms
of asymptotic expansion of the function 
$\omega ^\mathrm{sc}(\la,\mu|\kappa,\kappa,\al)$
are presented.

\section{Scaling limit in homogeneous and inhomogeneous cases}\label{lattice}

Our study of continuous models is based on the scaling limit of the homogeneous or the inhomogeneous
six vertex models on the cylinder. 
In this section we repeat several definitions and correct some inaccuracies
committed in \cite{HGSIV,OP}. We start with the infinite tensor product of $\mathbb{C}^2$
which is denoted by $\mathfrak{H}_\mathbf{S}$. 
In the inhomogeneous case we have alternating parameters $\z _0^{\pm2}$ attached to
every component of the tensor product. 
We define the operator $S(0)=
\half\sum_{j=-\infty}^0\sigma ^3_j$. Then the vector space $\mathcal{W}_{\al-s,s}$ consists
of the operators $q^{2(\al -s)S(0)}\mathcal{O}^{(s)}$ with $\mathcal{O}^{(s)}$ being local and of spin $s$.
All the operators which we consider later act on the space
$$\mathcal{W}^{(\al)}
=\bigoplus\limits _{s=-\infty}^{\infty}\mathcal{W}_{\al-s,s}
\,.
$$

In \cite{HGSII} we have defined
the creation operators $\tb^*(\z)$, $\bb_{\mathrm{rat}}^*(\z)$, $\cb_{\mathrm{rat}}^*(\z)$
and the annihilation operators $\bb (\z)$, $\cb (\z)$, all acting on $\mathcal{W}^{(\al)}$. 
We follow the notation
of  \cite{HGSIV}. Actually, the definition of these
operators is different in the homogeneous and the inhomogeneous cases, but using
formulae from \cite{HGSII} one can easily figure out what they are. We do not make distinction between the
homogeneous and the inhomogeneous cases notationally, all the explanations will be given at proper
places. 

The creation operators are not defined 
uniquely. One can 
apply Bogolubov transformations as we
shall discuss soon. Since we shall have a variety of different possibilities, in which 
one can be easily lost,  it is
important to start from the safe ground provided by the operators  $\z^{-\al}\bb_{\mathrm{rat}}^*(\z)$, $\z^{\al}\cb_{\mathrm{rat}}^*(\z)$. They are defined uniquely 
by the requirement:
$$\Tr_\mathbf{S}(\bb ^*_\mathrm{rat}(\z)(X))=0, \quad \Tr_\mathbf{S}(\cb ^*_\mathrm{rat}(\z)(X))=0,
\quad \forall X\in \mathcal{W}^{(\al)}\,.$$
Another important property is that $\z ^{-\al}\bb_{\mathrm{rat}}^*(\z)(X)$ and $\z ^{\al}
\cb_{\mathrm{rat}}^*(\z)(X)$ are rational functions of $\z ^2$.

The operator $\tb^*(\z)$ lies in the centre,
so, we shall handle it as a $\mathbb{C}$-number. Actually, for different reasons
we shall aim at having $\tb^*(\z)=2$ in our final formulae. This explains that the formulae
in \cite{OP} were written in the quotient space $\mathcal{W}^{(\al)}/(\tb^*(\z)-2)\mathcal{W}^{(\al)}$.
However, it is not quite consistent to work in the quotient space from the very beginning,
so, for the time being we keep $\tb ^*(\z)$.

Now following \cite{HGSIII,HGSIV} we define the operators 
$\bb ^*(\z)$, $\cb ^*(\z)$ by the Bogolubov transformation. 
\begin{align}
\bb^*(\z)=e^{-\widetilde{\Omega}}\ \bb  _\mathrm{rat}^*(\z)\ e^{\widetilde{\Omega}},\quad
\cb ^*(\z)=e^{-\widetilde{\Omega}}\ \cb  _\mathrm{rat}^*(\z)\ e^{\widetilde{\Omega}}\,,\label{modtbc}
\end{align}
where
\begin{align}
\widetilde{\Omega}=\frac 1 {(2\pi i)^2}\oint\limits_{\Gamma}\oint\limits_{\Gamma}
\tilde{\omega}(\z,\xi|\al)\cb (\xi)\bb (\z)\ \frac {d\z^2}{\z ^2}\frac {d\xi^2}{\xi ^2}\,,
\label{Oo}
\end{align}
and
$$\tilde{\omega}(\z,\xi|\al)=-\left[\Delta _\z +\tb ^*(\xi)-\tb^*(\z)
   + 4\delta ^-_\z \delta ^-_\xi\Delta ^{-1}_\z\right]\psi (\z/\xi,\al)\,,$$
we denote by
$\delta^-_\z$ and $\Delta _\z$  the following finite difference operators \begin{align}
&\delta^-_\z f(\z)=f(\z q)-\half \tb ^*(\z)f(\z )\,,\nn\\
&\Delta _\z f(\z)=f (\z q)-f(\z q^{-1})\,.\nn
\end{align}
We set
$$\psi (\z)=\z ^{\al}\frac {\z^2+1}{2(\z ^2-1)}\,.$$
We do not explain how the transcendental function $
\delta ^-_\z \delta ^-_\xi\Delta ^{-1}_\z\psi (\z/\xi,\al)$ is 
%understood (see \cite{HGSIV}), later 
understood (see \cite{HGSIV}). Later 
we shall write an
explicit formula in a particular case needed in this paper. 
The importance of the operators $\bb ^*(\z)$, $\cb ^*(\z)$ clearly follows from \cite{HGSIII,HGSIV},
we shall comment on it later. 

%The operators $\bb ^*(\z)$, $\cb ^*(\z)$ satisfy equations similar to \eqref{mainb}, \eqref{mainc} with
%new function 
%$$\ob_{\mathbf{n}}(\z,\xi|\kappa,\al)=\ob_\mathbf{\mathrm{rat},{n}}(\z,\xi|\kappa,\al)+
%\tilde{\omega}(\z,\xi|\al)\,.$$

Using the formulae of the subsection 2.5 in \cite{HGSII} one concludes that
\begin{align}
\z ^{-\al}\bb ^*(\z)\to 0, \quad
\z ^{\al}\cb ^*(\z)\to 0,\qquad \z \to 0\,.\label{goesto0}\end{align}
However, this property does not hold for $\bb_{\mathrm{rat}}^*(\z)$, $\cb_{\mathrm{rat}}^*(\z)$
because the function $\tilde{\omega}(\z,\xi|\al)$ spoils it. It is impossible to
define operators which are rational and satisfy \eqref{goesto0}. We shall define 
the operator $\bb ^*_0(\z)$ which 
satisfies {``}one half" of \eqref{goesto0}, i.e., the latter half only,
\begin{align}
\bb_0^*(\z)=e^{-\nabla\Omega}\ \bb  _\mathrm{rat}^*(\z)\ e^{\nabla\Omega},\quad
\cb _0^*(\z)=e^{-\nabla\Omega}\ \cb  _\mathrm{rat}^*(\z)\ e^{\nabla\Omega}\,,\label{nabla}
\end{align}
where  the operator
$\nabla\Omega$ is constructed as in \eqref{Oo} %from
replacing 
$\tilde{\omega}(\z,\xi|\al)$ by
\begin{align}
\nabla\omega(\z,\xi|\al)&=\left[\tb^*(\z)-\tb ^*(\xi)-\Delta _\z 
   \right]\psi (\z/\xi,\al)\nn \\ &-
 \frac{(2q^{\al}-\tb ^*(\z))(2q^{-\al}-\tb ^*(\xi))}{2(q^{\al}-q^{-\al})}
   \(\frac\z \xi\)^{\al}\,.
   \nn
   \end{align}
Combining \eqref{modtbc} with \eqref{nabla}   we obtain
\begin{align}
\bb^*(\z)=e^{-\Omega_0}\ \bb  _0^*(\z)\ e^{\Omega_0},\quad
\cb ^*(\z)=e^{-\Omega_0}\ \cb  _0^*(\z)\ e^{\Omega_0}\,,\label{bcb0c0}
\end{align}
where $\Omega_0$ given by \eqref{Oo} with $\tilde{\omega}(\z,\xi|\al)$ replaced by
$4\omega_0(\z,\xi|\al)$, and
\begin{align}
\omega_0(\z,\xi|\al)=
 -  \delta ^-_\z \delta ^-_\xi\Delta ^{-1}_\z\psi _0 (\z/\xi,\al),\quad%\nn\\
   \psi _0(\z,\al)=
   \frac {{\z}^{\al}}{\z ^2-1}\,.
\label{omega}
\end{align}
We note that
$$\tilde{\omega}(\z,\xi|\al)=\nabla\omega(\z,\xi|\al)+4\omega_0(\z,\xi|\al).$$
Now it is easy to see that since we are working in
$$0<\al<2\,,$$
the following estimates hold,
\begin{align}
&\bb ^*_0(\z)=O(\z ^{\al}),\quad \ \ \cb ^*_0(\z)=O(\z ^{2-\al}),\quad \z\to 0\,.\nn
\end{align}

The analytical structure of the  creation operators depends on the target space.
If the latter is $\mathcal{W}_{\al,0}$ we have for $\bb^*_0(\z)$, $\cb^*_0(\z)$:
\begin{align}
\bb _0^*(\z)=\sum\limits _{j=1}^{\infty}\z ^{\al -2+2j}\bb^*_{\mathrm{screen},j},\quad
\cb _0^*(\z)=\sum\limits _{j=1}^{\infty}\z ^{-\al+2j}\cb^*_{\mathrm{screen},j}\,,
\label{bcscreen}
\end{align}
the meaning of the suffix {``}screen" will be clear from what follows. 

Now let us return to the main problem solved in \cite{HGSIII}. 
We start with the six-vertex model on an
infinite cylinder. We call the infinite direction the {``}space direction", 
and the compact direction the {``}Matsubara direction".
We associate with them the spaces $\mathfrak{H}_\mathbf{S}$, $\mathfrak{H}_\mathbf{M}$.
The number of sites in the Matsubara direction is even, and is denoted by $\mathbf{n}$. 
In the present paper we shall consider two cases:
homogeneous six vertex model which gives the chiral CFT in the scaling limit \cite{HGSIV}, and
inhomogeneous six vertex model which gives the sG model in the scaling limit. In both cases we 
denote by $T_{\mathbf{S}, \mathbf{M}}$ the rectangular monodromy matrix representing
the universal $R$-matrix in the tensor product $\mathfrak{H}_\mathbf{S}\otimes\mathfrak{H}_\mathbf{M}$:
\begin{align}
&T_{j,\mathbf{M}}=\raisebox{.7cm}{$\curvearrowright $} 
\hskip -.75cm\prod\limits_{j=-\infty}^{\infty}
T_{j,\mathbf{M}} (1),
\qquad\ \ \ T_{j,\mathbf{M}}(\z)=\raisebox{.7cm}{$\curvearrowleft $} 
\hskip -.6cm\prod\limits_{\mathbf{m=1}}^{\mathbf{n}}
L_{j,\mathbf{m}}(\z), \qquad\ \ \ \ \  \mathrm{homogeneous,}\label{Thom}\\
&T_{\mathbf{S},\mathbf{M}}=\raisebox{.7cm}{$\curvearrowright $} 
\hskip -.75cm\prod\limits_{j=-\infty}^{\infty}
T_{j,\mathbf{M}} (\z_0 ^{(-1)^j})\,,\quad
T_{j,\mathbf{M}}(\z) =\raisebox{.7cm}{$\curvearrowleft $} 
\hskip -.6cm\prod\limits_{\mathbf{m=1}}^{\mathbf{n}}
L_{j,\mathbf{m}}(\z\z_0^{-(-1)^\mathbf{m}})
\,,\ \mathrm{inhomogeneous}.\label{Tinhom}
\end{align}

In the paper \cite{HGSIII} we considered the functional
\begin{align}
Z^{\kappa}_\mathbf{n}\Bigl\{q^{2\al S(0)}\mathcal{O}
\Bigr\}=
\frac{\Tr _{\mathrm{S}}\Tr _{\mathbf{M}}\Bigl(T_{\mathrm{S},\mathbf{M}}q^{2\kappa S+2\al S(0)}\mathcal{O}\Bigr)}
{\Tr _{\mathrm{S}}\Tr _\mathbf{M}\Bigl(T_{\mathrm{S},\mathbf{M}}q^{2\kappa S+2\al S(0)}\Bigr)}
\,.
\label{Z}
\end{align}

The main theorem of \cite{HGSIII}
says that
\begin{align}
&Z_\mathbf{n}^{\kappa}\bigl\{\tb^*(\z)(X)\bigr\}
=2\rho_\mathbf{n}(\z|\kappa,\kappa+\al)Z_\mathbf{n}^{\kappa}\{X\}\,,\label{maint}\\
&Z_\mathbf{n}^{\kappa}\bigl\{\bb^*(\z)(X)\bigr\}
=\frac 1{2\pi i}\oint\limits _{\Gamma}
\ob_\mathbf{n}(\z,\xi|\kappa,\al)
Z_\mathbf{n}^{\kappa}\bigl\{\cb(\xi)(X)\bigr\}
\frac{d\xi^2}{\xi^2}\,,\label{mainb}\\
&Z_\mathbf{n}^{\kappa}\bigl\{\cb^*(\z)(X)\bigr\}
=-\frac 1 {2\pi i}\oint\limits_{\Gamma}
\ob _\mathbf{n}(\xi,\z|\kappa,\al)
Z_\mathbf{n}^{\kappa}\bigl\{\bb(\xi)(X)\bigr\}
\frac{d\xi^2}{\xi^2}\,,
\label{mainc}
\end{align}
where the integration contour goes around $1$ for the homogeneous case and around $\z _0^{\pm 2}$
for the inhomogeneous one. We use the 
boldface
%bold face 
letter for $\ob _\mathbf{n}(\xi,\z|\kappa,\al)$ in order to distinguish it from several auxiliary $\omega$'s which we had before. 

%The distinguished character of $\bb ^*(\z)$, $\cb ^*(\z)$ 
%consists in the fact that for 
A distinguished feature of $\bb ^*(\z)$, $\cb ^*(\z)$ is that in
the simple-minded limit $\mathbf{n}\to \infty$ we have
$$Z_{\infty}^{\kappa}\bigl\{\bb^*(\z)(X)\bigr\}=0, \quad Z_{\infty}^{\kappa}\bigl\{\cb^*(\z)(X)\bigr\}=0\,.$$
The word {``}simple-minded" means that $\mathbf{n}$ goes to $ \infty$ without rescaling the
Bethe roots %of 
for
the Matsubara transfer-matrix. 

Certainly, the same kind of relations is true if we put $\bb ^*_0$ and $\cb^*_0$ 
in the left hand sides, and replace $\ob _\mathbf{n}$ by $\ob _\mathbf{n}+4\om _0$ in the right hand sides. 
We have
$$
\rho _\mathbf{n}(\z|\kappa,\kappa+\al)=\frac{T_\mathbf{n}(\z |\al+\kappa)}
{T_\mathbf{n}(\z |\kappa)}\,,
$$
where $T_\mathbf{n}(\z |\al+\kappa)$, $
T_\mathbf{n}(\z |\kappa)$ are maximal eigenvalues of the twisted Matsubara transfer-matrices \cite{HGSIII}. We do not write the equation for $\ob_\mathbf{n}(\z,\xi|\kappa,\al)
$ here. It can be found in \cite{HGSIII,BG,HGSIV}. 

Following \cite{HGSIV} we introduce a generalised functional 
%$Z^{\kappa,s}$, 
$Z^{\kappa,-s}_{\mathbf{n}}$, 
then we shall comment
on its relevance 
%for 
to
the scaling limit in the homogeneous and the inhomogeneous cases. 
If the target of the operator $\z^{\al}\cb ^*_0(\z)$ is $\mathcal{W}_{\al+s,-s}$ it develops a singularity at 
$\z^2=0$, namely,
\begin{align}
\cb _0^*(\z)=\sum\limits _{j=0}^{s-1}\z ^{-\al-2j}\cb^*_{\mathrm{screen},-j}+\cb _{0,\mathrm{reg}}^*(\z)\,,
\quad
\cb _{0,\mathrm{reg}}^*(\z)=\sum\limits _{j=1}^{\infty}\z ^{-\al+2j}\cb^*_{\mathrm{screen},j}\,.
\label{c0sing}
\end{align}
Using the singular part of $\z^{\al+2}\cb ^*_0(\z)$ we define:
\begin{align}
%Z^{\kappa,-s}
Z^{\kappa,-s}_{\mathbf{n}}
\Bigl\{q^{2\al S(0)}\mathcal{O}
\Bigr\}=
\frac{\Tr _{\mathrm{S}}\Tr _{\mathbf{M}}\Bigl(Y^{(s)}_\mathbf{M}T_{\mathrm{S},\mathbf{M}}
q^{2\kappa S}\ \cb ^*_{\mathrm{screen},-0}\cdots \cb ^*_{\mathrm{screen},-s+1}(
q^{2\al S(0)}\mathcal{O})\Bigr)}
{\Tr _{\mathrm{S}}\Tr _\mathbf{M}\Bigl(Y^{(s)}_\mathbf{M}T_{\mathrm{S},\mathbf{M}}
q^{2\kappa S}
\ \cb ^*_{\mathrm{screen},-0}\cdots \cb ^*_{\mathrm{screen},-s+1}(q^{2\al S(0)})\Bigr)}
\,,
\label{Zs}
\end{align}
where $Y^{(s)}_\mathbf{M}$ is an operator of spin $s$ acting only in $\mathfrak{H}_\mathbf{M}$.
This functional possesses several nice properties. First, it is independent of 
$Y^{(s)}_\mathbf{M}$ if the latter is  in general position \cite{HGSIV}. Second, 
%it allows the same computation 
%(\ref{maint}, \ref{mainb}, \ref{mainc}), only 
formulae \eqref{maint}, \eqref{mainb}, \eqref{mainc}
remain valid provided
the functions 
$\rho _\mathbf{n}(\z|\kappa+\al,\kappa)$ and $\ob_\mathbf{n}(\z,\xi|\kappa,\al)$ 
are replaced by appropriate counterparts.
%change. 
In particular,
\begin{align}
\rho _\mathbf{n}(\z|\kappa,\al,-s)= \frac{T_\mathbf{n}(\z |\al+\kappa+s,-s)}
{T_\mathbf{n}(\z |\kappa)}\,,
\end{align}
where $T_\mathbf{n}(\z |\al+\kappa+s,-s)$ is the maximal eigenvalue of the Matsubara transfer-matrix
with twist $\al+\kappa+s$ in the space of spin $-s$. 
For later reference we record also
\begin{align}
&Z^{\kappa,-s}_{\mathbf{n}}
\Bigl\{
\bb^*(\z)\cb^*(\xi)\bigl(q^{2\al S(0)}\bigr)
\Bigr\}=\ob_\mathbf{n}(\z,\xi|\kappa,\al,-s)
\,,
\label{omegan1}
\\
&
Z^{\kappa,-s}_{\mathbf{n}}
\Bigl\{
\bb_0^*(\z)\cb_0^*(\xi)\bigl(q^{2\al S(0)}\bigr)
\Bigr\}
=\ob_\mathbf{n}(\z,\xi|\kappa,\al,-s)+4\omega_0(\z,\xi|\al)
\,.\label{omegan2}
\end{align}

Now we shall discuss the importance
of %this definition 
the definition \eqref{Zs}
for the scaling limit in the homogeneous and the inhomogeneous cases.
\vskip .3cm
\noindent
{\bf Homogeneous case.} Recall \cite{HGSIV} that introducing the step of the lattice $a$ we 
consider for the homogeneous chain
the scaling limit 
\begin{align}
\mathbf{n}\to\infty, \quad a\to 0, \quad \mathbf{n}  a=2\pi \ \ \mathrm{fixed}\,.
\end{align}
Actually in \cite{HGSIV} we wrote $2\pi R$ in the right hand side of the last formula,
but in this paper we prefer to set $R=1$ in order to avoid confusion with the sG case.
Anyway, in the conformal case the dependance on the radius can be easily reconstituted. 
The spectral parameter $\z$ is subject to rescaling 
\begin{align} \z = \la (C a)^{\nu}\,, \label{zetala}\end{align}
where $\la$ is finite and
\begin{align}
C=\  \frac{\Gamma\(\frac {1-\nu}{2\nu}\)}{2\sqrt{\pi}\ \Gamma \(\frac 1{2\nu}\)}\Gamma (\nu)^{\frac 1\nu}\,.
\nn
\end{align}
We have
$$
\lim_{a\to 0}
\rho _\mathbf{n}(\z|\kappa,\al,-s)=
%\ \ \longrightarrow
%\hskip -1cm {}_{{\ }_{\mathrm{scaling}}}
\ \ \rho (\la|\kappa,\kappa')=\frac{T^\mathrm{sc}(\la |\kappa ')}
{T^\mathrm{sc}(\la |\kappa)}\,, \quad \kappa '=\kappa +\al -2s\ {\textstyle \frac{1-\nu}\nu}\,,
$$
where $T^\mathrm{sc}(\la |\kappa)$ is the maximal eigenvalue of the BLZ transfer-matrix
\cite{BLZI,BLZII}. By analytical continuation we allow $\kappa '$ to be arbitrary.

The main statement of  \cite{HGSIV} is that the functional
$Z^{\kappa,-s}$ scales to the normalised three point function in CFT.
Namely if a quasi-local operator $q^{2\al S(0)}\mathcal{O}$ tends to
a Virasoro descendant
$P_{\al,\mathcal{O}}(\{\mathbf{l}_{-k}\})\Phi _{\al}(0)$
with $P_{\al,\mathcal{O}}$ a polynomial, then
\begin{align}
\lim_{a\to0}
Z^{\kappa,-s}\Bigl\{q^{2\al S(0)}\mathcal{O}
\Bigr\}
=
%\ \ \longrightarrow
%\hskip -1cm {}_{{\ }_{\mathrm{scaling}}} \ \ 
\frac {\langle 1-\kappa '|P_{\al,\mathcal{O}}(\{\mathbf{l}_{-k}\})\Phi _
{\al}(0)|1+\kappa\rangle}
{\langle 1-\kappa '|\Phi _{\al}(0)|1+\kappa\rangle}\,,
\label{Zks1}
\end{align}
where the right hand side is a ratio of three-point
functions in CFT with central charge
$$c=1-6\frac {\nu ^2}{1-\nu}\,.$$
For technical reasons, we have been able to treat quantitatively only the case $\kappa=\kappa '$ \cite{HGSIV}.
In that case 
\begin{align}\rho (\la|\kappa,\kappa)=1\,,\label{rho=1}\end{align}
and we can factor out the action of $\tb ^*(\z)$ since \eqref{rho=1} implies
\begin{align}
\lim_{a\to0}\tb^*(\z)=
%\ \ \longrightarrow
%\hskip -1cm {}_{{\ }_{\mathrm{scaling}}}\ \ 
2\,.\label{t=2}
\end{align}
So, this important condition appears in CFT case as a technical requirement. 
\vskip .3cm
\noindent{\bf Inhomogeneous case.} In this case we follow \cite{DDV}. The main idea is
to introduce again the step of the lattice $a$ and radius $R$ related by
$$\mathbf{n}a=2\pi R\,,$$
and to consider the limit
$$a\to 0,\ \ \z_0\to \infty, \qquad \mathrm{keeping}\ R\ \mathrm{and}\ M=4a^{-1}\z_0^{-\frac 1 \nu} \ \ \mathrm{fixed}\,.$$
Then we are supposed to obtain the Euclidean sG model on a cylinder of radius $R$
with the mass of soliton equal to $M$. We want to consider 
$$\frac{\langle P(\{\mathbf{l}_{-k}\},\{\bar{\mathbf{l}}_{-k}\})\Phi _{\al}(0) \rangle ^{\mathrm{sG}}_R}
{\langle  \Phi _{\al}(0) \rangle ^{\mathrm{sG}}_R}\,.$$
%For irrational $\nu$,
the descendants are uniquely defined by their conformal limit and the absence 
of finite counterterms which can be guaranteed  by dimensional reasons. 
By dimensional reasons,
for irrational $\nu$, the descendants are uniquely defined by their conformal limit 
and the absence of finite counterterms.

The first na\"ive idea would be to consider the scaling limit of $Z^{0}_\mathbf{n}$. We set
$\kappa =0$ in order to simplify the formulae, though one can easily generalise. However,
this idea immediately proves to be wrong because of our old enemy $\tb ^*(\z)$.
Indeed, according to \eqref{maint} it gives a non-trivial contribution
to $Z^{0}_\mathbf{n}$. But we know that $\tb ^*(\z)$ is the generating function of the
adjoint action of the local integrals of motion. So, nontrivial $\rho _\mathbf{n}(\z|\al,0)$
breaks this invariance, in particular, it breaks the translational invariance. 

What is the reason for that? The point is that due to the bosonisation rule
$$
\lim_{a\to0}a^{-1}\
\sigma ^3_j= \frac {1}{i\pi (1-\nu)}\partial _x\varphi (x)\,,
%\ \ \longrightarrow
%\hskip -1cm {}_{{\ }_{\mathrm{scaling}}}
%\ \ \frac {a}{i\pi (1-\nu)}\partial _x\varphi (x)\,,
$$
we have
$$
\lim_{a\to0}
q^{2\al S(0)}=\Phi _{-\al}(-\infty)\Phi _{\al}(0)\,.
%\ \ \longrightarrow
%\hskip -1cm {}_{{\ }_{\mathrm{scaling}}}\ \ 
%\Phi _{\al}(0)\Phi _{-\al}(-\infty)\,.
$$
So, the scaling limit of $Z_{\mathbf{n}}\{q^{2\al S(0)}\mathcal{O} \}$ corresponds rather to the
two-point function with one field placed at $-\infty$ than to the one-point function.
There is no simple way to define an analogue of the field $\Phi _{\al}(0)$ itself on the
lattice. So, we have to take an indirect way in order to "screen out" the field $\Phi _{-\al}(-\infty)$,
and to restore the invariance under the action of the local integrals of motion.

It is true that the difficulty is the same as in the previous case, 
but the implications are different. 
While the requirement \eqref{t=2} appears as a technical restriction 
%on the choice of $\kappa$ 
in the CFT case, 
it is of direct physical significance in the sG model. 
We satisfy this requirement in the same way as in CFT assuming that 
\begin{align}
\lim_{a\to0}
Z_\mathbf{n}^{-s}\{q^{2\al S(0)}\mathcal{O}\}=
%\ \ \longrightarrow
%\hskip -1cm {}_{{\ }_{\mathrm{scaling}}}\ \ 
\frac{\langle \ 
P_{\al,\mathcal{O}}(\{\mathbf{l}_{-k}\},\{\bar{\mathbf{l}}_{-k}\})
\Phi _{\al}(0)\ \rangle ^{\mathrm{sG}}_R}
{\langle \ \Phi _{\al}(0)\ \rangle ^{\mathrm{sG}}_R}
%{\Tr \(e^{-2\pi RH ^\mathrm{sG}}P_{\mathcal{O}}(\{\mathbf{l}_{-k}\},\{\bar{\mathbf{l}}_{-k}\})
%\Phi _{\al}(0)\)}
%{\Tr \(e^{-2\pi RH ^\mathrm{sG}}\Phi _{\al}(0)\)}
\,,\quad \al=2s{\textstyle\frac {1-\nu}\nu}\,.\label{mainconj}
\end{align}
Certainly the statements \eqref{Zks1}, 
\eqref{mainconj} are rather strong conjectures.  
We would be happy to have more robust supporting arguments than we have for the moment.
Notice that the analytical continuation from the
points $\al=2s{\textstyle\frac {1-\nu}\nu}$ is possible.

%%%%%%%%%%%%%%%%%%%%%%%%%%%%%

\section{Fermions in chiral CFT}

In this section we study the fermions and their expectation values 
in chiral CFT.
Recall %\cite{OP} 
that %doing 
performing the scaling limit we rescale
$\z$ according to \eqref{zetala}.
In the scaling limit, 
%the operators 
$\bb ^*(\z)$, $\cb ^*(\z)$ produce %the 
operators $\betab^*(\la)$, 
$\gammab ^*(\la)$ with the asymptotics at $\la =\infty$:
\begin{align}
&\betab ^*(\la)=\half\lim\limits_{a\to 0}\bb ^*(\la(Ca)^{\nu})  ,\quad
\gammab ^*(\la)=\half\lim\limits_{a\to 0}\cb ^*(\la(Ca)^{\nu})\,,\nn\\
&\betab^*(\la)\simeq\sum\limits _{j=1}^{\infty}\la ^{-\frac{2j-1}\nu}\betab^*_{2j-1},\quad
\gammab^*(\la)\simeq\sum\limits _{j=1}^{\infty}\la ^{-\frac{2j-1}\nu}\gammab^*_{2j-1}\,,\label{betgam}
\end{align}
while $\bb ^*_0(\z)$, $\cb ^*_0(\z)$ produce operators 
$\betab _{\mathrm{screen}}^*(\la)$, 
$\gammab_{\mathrm{screen}}^*(\la)$ with the asymptotics at $\la=0$:
\begin{align}
&\betab_\mathrm{screen} ^*(\la)=\half\lim\limits_{a\to 0}\bb _{0}^*(\la(Ca)^{\nu})\quad
\gammab_\mathrm{screen} ^*(\la)=\half\lim\limits_{a\to 0}\cb _{0}^*(\la(Ca)^{\nu}),\nn\\
&\betab _{\mathrm{screen}}^*(\la)\simeq\sum\limits _{j=1}^{\infty}\la ^{\al+2j-2}
\betab_{\mathrm{screen},j}^*\,,\quad
\gammab_{\mathrm{screen}}^*(\la)\simeq
\sum\limits _{j=1}^{\infty}\la ^{-\al+2j}\gammab_{\mathrm{screen},j} ^*\,.\label{scr1}
\end{align}
We assign scaling dimensions to
these operators by the corresponding powers in $\la$. % in their expansions \eqref{betgam,scr1}.
Namely $\betab^*_{2j-1}$,  $\gammab^*_{2j-1}$ carry the scaling dimension $2j-1$, 
while $\betab_{\mathrm{screen},j}^*$ and $\gammab^*_{\mathrm{screen},j}$ 
carry the scaling dimensions $\nu(2-\al-2j)$ and 
$\nu(\al-2j)$, respectively.

%The fermionic picture implies that the 
The
normalised three-point functions
$$
\frac {\langle 1-\kappa|\betab ^*_{I^+}\gammab^*_{I^-}
\betab ^*_{\mathrm{screen},J^+}\gammab ^*_{\mathrm{screen},J^-}\phi _{\al}(0)|1+\kappa\rangle}
{\langle 1-\kappa|\phi _{\al}(0)|1+\kappa\rangle}
$$
can be expressed as determinants of pairings.
% for which we first give formulae 
%and then explain the notation and the origin. 
Each pairing is 
written in terms of  
a function $\Theta(l,m|\kappa,\al)$  introduced in \cite{HGSIV}.
Its asymptotics for $\kappa\to \infty$
can be computed by a regular procedure. 
For reference, 
in  Appendix \ref{app2} we provide the asymptotic expansion
up to the order $\kappa ^{-6}$.
Motivated by this asymptotic expansion we conjecture that
$\Theta(l,m|\kappa,\al)
+i/(l+m)$ 
is an entire function of $l,m$.

Set \begin{align}
&D_{2j-1}(\al)=-\sqrt{\frac i \nu}\ \Gamma (\nu)^{-\frac {2j-1}\nu}(1-\nu)^{\frac{2j-1} 2}
\frac{\Gamma \(\frac \al 2 +\frac 1 {2\nu}(2j-1)\)}{(j-1)!\Gamma \(\frac \al 2 +\frac {1-\nu} {2\nu}(2j-1)\)}\,,\label{DE}\\
&E_j(\al)=\frac{(-1)^{j-1}}{\sqrt{i}}\ \( 2i \Gamma(\nu)^{\frac 1 \nu}  \nu^{-1}\)^{(2j-\al)\nu}
\frac{\Gamma (\frac 1 2 +\nu j -\frac {\nu\al} 2)}{(j-1)!\Gamma (1-(1-\nu)j-\frac {\nu\al}2)}\,.\nn
\end{align}
Notation being as above, we have:
%We have:
\begin{align}
&\frac {\langle 1-\kappa|\betab ^*_{2j-1}\gammab^*_{2k-1}\phi _{\al}(0)|1+\kappa\rangle}
{\langle 1-\kappa|\phi _{\al}(0)|1+\kappa\rangle}=\bigl({\textstyle\frac{\nu}{2\sqrt{1-\nu}}}\kappa\bigr)^{2(j+k)-2}
\nu ^{-1}\label{par1}\\
&\qquad\times D_{2j-1}(\al)D_{2k-1}(2-\al)\Theta(i{\textstyle\frac {2j-1}{2\nu}},i{\textstyle\frac {2k-1}{2\nu}}
|\kappa,\al)\,,\nn\\
&\frac {\langle 1-\kappa|\betab ^*_{\mathrm{screen},j}\gammab^*_{\mathrm{screen},k}
\phi _{\al}(0)|1+\kappa\rangle}{\langle 1-\kappa|\phi _{\al}(0)|1+\kappa\rangle}=\kappa^{-2\nu(j+k-1)}\label{par2}\\
&\qquad\times E_{j}(2-\al)E_{k}(\al)\Theta(-i(j-1+{\textstyle\frac \al 2}),-i(k-{\textstyle\frac \al 2})|\kappa,\al)\,,\nn\\
&\frac {\langle 1-\kappa|\betab ^*_{2j-1}\gammab^*_{\mathrm{screen},k}\phi _{\al}(0)|1+\kappa\rangle}
{\langle 1-\kappa|\phi _{\al}(0)|1+\kappa\rangle}=\bigl({\textstyle\frac{\nu}{2\sqrt{1-\nu}}}\kappa\bigr)^{2j-1}
\nu ^{-\frac 1 2}\kappa^{-2\nu k+\nu\al}\label{par3}\\
&\qquad\times D_{2j-1}(\al)E_{k}(\al)\Theta(i{\textstyle\frac {2j-1}{2\nu}},-i(k-{\textstyle\frac \al 2})|\kappa,\al)\,,\nn\\
&\frac {\langle 1-\kappa|\betab ^*_{\mathrm{screen},j}\gammab^*_{2k-1}\phi _{\al}(0)|1+\kappa\rangle}
{\langle 1-\kappa|\phi _{\al}(0)|1+\kappa\rangle}=\bigl({\textstyle\frac{\nu}{2\sqrt{1-\nu}}}
\kappa\bigr)^{2k-1}\nu ^{-\frac 1 2}\kappa^{-2\nu (j-1)-\nu \al}\label{par4}\\
&\qquad\times E_{j}(2-\al)D_{2k-1}(2-\al)\Theta(-i(j-1+{\textstyle\frac \al 2}),i{\textstyle\frac {2k-1}{2\nu}}
|\kappa,\al)\,.\nn
\end{align}
%where

Let us explain the origin of the formulae (\ref{par1})--(\ref{par4}). 

Formula  \eqref{par1} arises from the scaling limit of \eqref{omegan1}. The function
\begin{align*}
\omega ^\mathrm{sc}(\la,\mu|\kappa,\kappa,\al)&=
{\textstyle\frac 1 4}%\lim_{\mathrm{scaling}}
\lim_{a\to 0}
\ob_\mathbf{n}(\z,\xi|\kappa,\al,-s) 
\end{align*}
is given by the inverse
Mellin transform of the function $\Theta(l,m|\kappa,\al)$ \cite{HGSIV}:
\begin{align}
\omega ^\mathrm{sc}(\la,\mu|\kappa,\kappa,\al)&=\frac{1}{2\pi i}\int\!\!\int
dl\,dm\,\tilde{S}(l,\alpha)\tilde{S}(m,2-\alpha)\Theta(l+i0,m|\kappa,\alpha)\,\label{omegaconf}\\
&\qquad\qquad \qquad\quad\times\Bigl(\frac{e^{\frac{\pi i\nu}2}\Gamma (\nu)2^\nu\la}
{(\nu \kappa)^{\nu}}\Bigr)^{2il}\Bigl(\frac{e^{\frac{\pi i\nu}2}\Gamma (\nu)2^\nu\mu}
{(\nu \kappa)^{\nu}}\Bigr)^{2im}\,,\nn\\
\tilde{S}(k,\alpha)&=\frac{\Gamma\bigl(-ik+\frac{\alpha}{2}\bigr)\Gamma\bigl(\frac{1}{2}+i\nu k\bigr)}
{\sqrt{2\pi}\ \Gamma\bigl(-i(1-\nu)k+\frac{\alpha}{2}\bigr)(1-\nu)^{\frac {1-\al}2}}\,.\nn
\end{align}
We obtain \eqref{par1} by closing the contours of
integration into the upper half plane %in the last formula. 
and taking the coefficients of powers of $\lambda$, $\mu$.

Next consider the scaling limit of \eqref{omegan2}.
As was said in Section 2, 
we consider only the case $\kappa =\kappa '$. So, effectively $\tb^*(\z)=2$, and the
function $\omega _0 (\z,\xi|\al)$ depends only on the ratio $\z/\xi$,
%on $\z/\xi$ %only
\begin{align}
&\omega _0 (\z,\xi|\al)=\omega _0 (\z/\xi,\al)\,,\label{omz}\\
&\omega _0(\la,\al)=- i\int\limits _{-\infty }^{\infty }\la ^{2ik}\frac{\sinh\frac{\pi}2(2(1-\nu)k+i\al)}
{2\sinh\frac{\pi}2(2k+i\al)\cosh\pi\nu k }dk\,.
\nn
\end{align}
We have
\begin{align}
\omega_0(\la^{-1},\al)=\omega_0(\la,2-\al)\,.\label{OMEGA0SYM}
\end{align}
Now 
%notice that 
%\textcolor{blue}{due to the above conjecture
%about the analytical structure of $\Theta(l,m|\kappa,\alpha)$ 
we have
\begin{align}
&\omega ^\mathrm{sc}(\la,\mu|\kappa,\kappa,\al)+\omega_0(\la/\mu,\al)\label{om+om}\\
&=\frac{1}{2\pi i}\int\!\!\int dl\,dm\,\tilde{S}(l,\alpha)\tilde{S}(m,2-\alpha)\Theta(l-i0,m|\kappa,\alpha)\,\nn\\
& \qquad\qquad\qquad\quad\times\Bigl(\frac{e^{\frac{\pi i\nu}2}\Gamma (\nu)2^\nu\la}
{(\nu \kappa)^{\nu}}\Bigr)^{2il}\Bigl(\frac{e^{\frac{\pi i\nu}2}\Gamma (\nu)2^\nu\mu}
{(\nu \kappa)^{\nu}}\Bigr)^{2im}\,.\nn
\end{align}
Assuming the conjectured
analyticity of $\Theta(l,m|\kappa,\alpha)$ and 
closing the contours into the lower-half plane,
we obtain \eqref{par2}.

Consider now \eqref{par3}. 
The corresponding generating function is
$\betab^*(\la)\gammab ^*_{\mathrm{screen}}(\mu)$, which seems to pose
a problem because we need to close contours into different half-planes. 
However, the difference between
\eqref{om+om} and \eqref{omegaconf} is the function %$\omega_0(\la/\mu|\al)$ 
$\omega_0(\la/\mu, \al)$ 
which for $\la\to\infty$, $\mu\to 0$
is given by asymptotical series containing $(\la/\mu)^{-\frac{2j-1}\nu}$ and $(\la/\mu)^{\al-2j}$
(see \eqref{asymomega} below). So, this asymptotics does not mix $\la ^{-\frac{2j-1}\nu}$ with $\mu^{-\al+2j}$
and we can safely use residues of \eqref{omegaconf} to compute \eqref{par3}. 
Similarly we obtain \eqref{par4}.

Of course the entire working carries over to the second chirality, 
but we have to recalculate $\omega$. 
Carefully repeating the computations of \cite{HGSIV} we get
\begin{align}
&
\overline{\omega}{\,}^{\mathrm{sc}}(\la,\mu|\kappa,\kappa,\alpha)
\simeq
\frac{1}{2\pi i}
\int\!\!\int dl dm
\tilde{S}(l,2-\alpha)
\tilde{S}(m,\alpha)
\Theta(l+i0,m|-\kappa,2-\alpha)
\label{omegabar}\\
&\qquad\qquad \qquad\qquad\qquad\quad\times
\Bigl(\frac{
e^{\frac{-\pi i\nu}2}\Gamma (\nu)2^\nu}{(\nu \kappa)^{\nu}\la }\Bigr)^{2il}
\Bigl(\frac{e^{\frac{-\pi i\nu}2}\Gamma (\nu)2^\nu}{(\nu \kappa)^{\nu}\mu}\Bigr)^{2im}
\,,
\nn\\
&\bar\omega_0(\la,\mu|\al)=\omega_0(\la/\mu,2-\al)\,.\nn
\end{align}
These functions are used to compute the pairings for the
operators
\begin{align}
&\bar{\betab}^*(\la)=\sum\limits _{j=1}^\infty\la ^{\frac{2j-1}\nu}
\bar{\betab}^*_{2j-1}\,,\quad
\bar{\gammab}^*(\la)=\sum\limits _{j=1}^\infty\la ^{\frac{2j-1}\nu}
\bar{\gammab}^*_{2j-1}\,,\label{BAREXPANSION1}\\
&\bar{\betab}_{\mathrm{screen}}^*(\la)
=\sum\limits _{j=1}^\infty\la ^{-2j+\al}\bar{\betab}^*_{\mathrm{screen}, j}\,,
\quad
\bar{\gammab}_{\mathrm{screen}}^*(\la)
=\sum\limits _{j=1}^\infty\la ^{-2j+2-\al}\bar{\gammab}^*_{\mathrm{screen}, j}\,.\label{BAREXPANSION2}
\end{align}

\section{Fermionic construction of primary fields}
\label{chiral}

Returning to the first chirality, we now 
explain how to use 
fermions to construct different primary fields out of one.

Acting on the primary field $\phi _\al (0)$,
the collection of operators
$\betab^*_{2j-1}$,  $\gammab^*_{2j-1}$, $\betab_{\mathrm{screen},j}^*$, $\gammab_{\mathrm{screen},j} ^*$
generates a huge space $\mathcal{H}_{\al}$. 
Unless stated otherwise, 
we shall always consider the quotient space modulo the action of the 
integrals of motion. 
We know from \cite{HGSIV} that the quotient space
$\mathcal{V}^\mathrm{quo}_{\al}$ is embedded into $\mathcal{H}_{\al}$ %this space 
with the basis being
$$\betab ^*_{I^+}\gammab^*_{I^-}\phi _{\al}(0)\,,$$
where $\#(I^+)=\#(I^-)$.
The main statement of the present paper concerning % the 
CFT is that the quotient space  $\mathcal{V}^\mathrm{quo}_{\al+2m\frac {1-\nu}\nu}$ can 
also be embedded into $\mathcal{H}_\al$, 
%be embedded in $\mathcal{H}_\al$ 
with the basis being
\begin{align}
\betab ^*_{I^+}\gammab^*_{I^-}\gammab ^*_{\mathrm{screen},I(m)}\phi _{\al}(0)\,,
\label{XXX}
\end{align}
with $\#(I^+)=\#(I^-)+m$. 
From the rule for assigning the scaling dimensions, 
one can easily 
conclude that the character of the space generated by \eqref{XXX} coincides with that of 
$\mathcal{V}^\mathrm{quo}_{\al+2m\frac {1-\nu}\nu}$. So, at least our statement makes sense from dimensional point of view.

In particular,
the vector of the lowest dimension among \eqref{XXX} must
be identified with the primary field:
\begin{align}
\phi _{\al +2m\frac {1-\nu}\nu}(0)\cong \betab^*_{\Io}
\gammab ^*_{\mathrm{screen},I(m)}\phi _{\al}(0)\,.\label{PRIMARY}
\end{align}
It has been said in Introduction that we use the symbol $\cong$ for identifying vectors belonging to different
spaces. For example, in the last formula the left hand side belongs
to $\mathcal{V}^\mathrm{quo}_{\al +2m\frac {1-\nu}\nu}$ while the right hand side belongs to  
$\mathcal{H}_\al$.
%The identification of these vectors goes through verifying that their three-point functions
%on the cylinder with the asymptotical conditions given by the primary fields
%$\phi _{1-\kappa}$, $\phi _{1+\kappa}$ coincide. 
We identify these vectors on the grounds that 
their three-point functions on the cylinder coincide,  
in the presence of primary fields 
$\phi _{1-\kappa}(-\infty)$, $\phi _{1+\kappa}(\infty)$ for arbitrary $\kappa$. 
Certainly, this is not enough to
state a theorem, but this is the best we can do for the moment. 

%For example, the 
The 
self-consistency of the formula \eqref{PRIMARY}
gives rise to an identity among the special values of
%functions 
%$\Theta$. 
$\Theta(l,m|\kappa,\alpha)$. 
There are two ways to
construct the field $\phi _{\al +4\frac{1-\nu}\nu}$ using fermions:
\begin{align}
\phi _{\al +4\frac{1-\nu}\nu}(0)&\cong\betab ^*_1\gammab ^*_{\mathrm{screen},1}\phi _{\al +2\frac{1-\nu}\nu}(0)\nn\\
&\cong\ \betab ^*_1\betab ^*_3\gammab ^*_{\mathrm{screen},2}\gammab ^*_{\mathrm{screen},1}\phi _{\al }(0)\,,\nn
\end{align}
where three fields belong correspondingly to
$\mathcal{V}^\mathrm{quo}_{\al +4\frac {1-\nu}\nu}$,  $\mathcal{H}_{\al +2\frac {1-\nu}\nu}$, $\mathcal{H}_{\al }$.
The creation operators $\betab ^*_1,\gammab ^*_{\mathrm{screen},1}$ in the first line and those in the second line
are different: the former acting on $\mathcal{H}_{\al +2\frac {1-\nu}\nu}$, and the latter on $\mathcal{H}_{\al }$.
This compatibility requirement implies the identity
\begin{align}
&\Theta (i/2\nu, -i(1-\al/2)|\kappa,\al)\Theta (i/2\nu, -i(1-\al/2-(1-\nu)/\nu)|
\kappa,\al +2(1-\nu)/{\nu})\nn\\&= -\frac 1 {4\nu}(\al\nu -2\nu +3)(\al\nu -4\nu +1)
\nn\\
&\times\left|\ \begin{matrix}\Theta (i/2\nu,-i(1-\al/2)|\kappa,\al)
&\Theta (i/2\nu,-i(2-\al/2)|\kappa,\al)\\
\Theta (3i/2\nu,-i(1-\al/2)|\kappa,\al)
&\Theta (3i/2\nu,-i(2-\al/2)|\kappa,\al)
\end{matrix}
\ \right|\,.\nn
\end{align}
Using the asymptotic expansion of $\Theta(l,m|\kappa,\alpha)$,
we have checked
%We check 
this identity up to $\kappa ^{-8}$. 
We regard 
the validity of the identity as another supporting evidence in favour of our 
statement \eqref{XXX}. 
Later in section section \ref{cons}, 
we shall give a proof of the corresponding identities in the more general 
setting of the sG model. 
\section{ Fermionic construction of descendants}

The descendants of $\phi _\al(0)$ can be constructed in the form %as
$$\betab ^*_{I^+}\gammab ^*_{I^-}\phi _\al(0)\,.$$
%We have the correspondence between the fermionic basis and 
They are related to the Virasoro descendants as
\begin{align}
&\betab ^*_{I^+}\gammab ^*_{I^-}\phi _\al(0)
\nn\\ &=\prod
\limits _{2j-1\in I^+}
D_{2j-1}(\al)\prod\limits _{2j-1\in I^-}
D_{2j-1}(2-\al)
\bigl[P_{I^+,I^-}^\mathrm{even}
%P_{I^+,I^-}^\mathrm{even}(\{\mathbf{l}_{-2k}\})
+d_{\al}
%P^\mathrm{odd}_{I^+,I^-}(\{\mathbf{l}_{-2k}\})
P^\mathrm{odd}_{I^+,I^-}
\bigr]\phi _\al(0)\,,\nn
\end{align}
%Let us give a few examples of relation between the fermionic basis and Virasoro descendants following \cite{HGSIV}:
%\begin{align}
%&\betab^*_1\gammab^*_1\phi _{\al}(0)=D_1(\al)D_1(2-\al)\mathbf{l}_{-2}\phi _{\al}(0)\,,\label{descen}\\
%&\betab^*_1\gammab^*_3\phi _{\al}(0)=\half D_1(\al)D_3(2-\al)\(\mathbf{l}_{-2}^2
%+\(\frac {2c-32} 9 +\frac 2 3 d_{\al}\)\mathbf{l}_{-4}\)\phi _{\al}(0)\,,\nn\\
%&\betab^*_3\gammab^*_1\phi _{\al}(0)=\half D_3(\al)D_1(2-\al)\(\mathbf{l}_{-2}^2
%+\(\frac {2c-32} 9 -\frac 2 3 d_{\al}\)\mathbf{l}_{-4}\)\phi _{\al}(0)\,,\nn
%\end{align}
where $D_{2j-1}(\al)$ is given in \eqref{DE}, and
\begin{align*}
&d_\alpha=\frac{\nu(\nu-2)}{\nu-1}(\alpha-1)=\textstyle{\frac 1 6}\sqrt{(25-c)(24\Delta _{\al}+1-c)}\,.
\end{align*}
As it was mentioned already, 
all formulas are to be understood modulo the action of the integrals of motion.
Here $P_{I^+,I^-}^\mathrm{even}$, $P_{I^+,I^-}^\mathrm{odd}$  
are polynomials in the generators $\{\mathbf{l}_{-2k}\}$, whose coefficients 
%The coefficients of the polynomials 
%$P_{I^+,I^-}^\mathrm{even}(\{\mathbf{l}_{-2k}\})$, 
%$P_{I^+,I^-}^\mathrm{odd}(\{\mathbf{l}_{-2k}\})$ 
depend polynomially on $c$ and 
rationally on $\Delta _{\al}$. 
The simplest examples are
\begin{align}
&%P_{\{1\},\{1\}}^\mathrm{even}(\{\mathbf{l}_{-2k}\})
P_{\{1\},\{1\}}^\mathrm{even}
=\mathbf{l}_{-2},
\quad
%P_{\{1\},\{1\}}^\mathrm{odd}(\{\mathbf{l}_{-2k}\})
P_{\{1\},\{1\}}^\mathrm{odd}
=0\,,\label{descen}\\
&%P_{\{1\},\{3\}}^\mathrm{even}(\{\mathbf{l}_{-2k}\})
P_{\{1\},\{3\}}^\mathrm{even}
=
%P_{\{3\},\{1\}}^\mathrm{even}(\{\mathbf{l}_{-2k}\})
P_{\{3\},\{1\}}^\mathrm{even}
=
\frac 1 2\mathbf{l}_{-2}^2
+\frac {2c-32} {18}\mathbf{l}_{-4}\,,\qquad
%\nn\\&
%P_{\{1\},\{3\}}^\mathrm{odd}(\{\mathbf{l}_{-2k}\})
P_{\{1\},\{3\}}^\mathrm{odd}
=
%P_{\{3\},\{1\}}^\mathrm{odd}(\{\mathbf{l}_{-2k}\})
P_{\{3\},\{1\}}^\mathrm{odd}
=-\frac 1 3\mathbf{l}_{-4}\,.\nn
\end{align}

The descendant $\betab ^*_{I^+}\gammab ^*_{I^-}\phi _{\al+2m\frac {1-\nu}\nu}(0)$
of the shifted primary field belongs to $\mathcal{V}^\mathrm{quo}_{\al +2m\frac {1-\nu}\nu}$.
We want to write another representation for this descendant using our definition of the
primary field $\phi _{\al+2m\frac {1-\nu}\nu}(0)$ in $\mathcal{H}_\al$.
Some general considerations
and, most importantly, \ ``experimental" data bring us to the following formula:
\begin{align}
\betab ^*_{I^+}\gammab ^*_{I^-}\phi _{\al+2m\frac {1-\nu}\nu}(0)
\cong \betab ^*_{I^++2m}\gammab ^*_{I^--2m}\ \betab^*_{\Io}\gammab^*_{\mathrm{screen},I(m)}
\phi _{\al}(0)\,.\label{idendesc}
\end{align}
There is a trouble here: 
it is obscure
%we do not know  how to understand $\gammab^*_{2j-1}$ 
%with negative $2j-1$ which may appear. 
how to understand $\gammab^*_{-a}$ 
when the suffix $-a$ becomes negative.
%The 
A
natural idea would be to %give it some meaning by using 
identify it with the annihilation operator
$\betab_a$ satisfying
\begin{align}
[\betab _a,\betab ^*_b]_+=\delta_{a,b}\varepsilon(a).\label{PAIRING}
\end{align}
These operators should originate from $\bb (\z)$. However, we do not know how to take the
continuous limit directly, and normalise these operators. So, we just introduce them by hand, and impose the rule
%The transformation rule is such that 
%if $\gammab^*_b$ in the left hand side of \eqref{idendesc}
%is such that $b<2m$ it is transformed into
\begin{align}
\gammab^*_{b-2m} =\betab_{2m-b} \quad \text{ if $b<2m$} 
%\gammab^*_b \rightarrow c(b)\betab_{2m-b}
\label{TRANS}
\end{align}
in the right hand side of \eqref{idendesc}. 
%Then, the 
The coefficient $\varepsilon(a)$ in \eqref{PAIRING}
is determined from the self-consistency, namely, they are
\begin{align}
& \varepsilon(a)=\textstyle{\frac 1 {i\nu}}\cot\textstyle{\frac\pi{2\nu}}(\nu\al+a).
\label{conjcomm}
\end{align}
Let us see why this prescription is good. 
Consider the formula \eqref{idendesc} together with several identities. 
 
First,
\begin{align*}
&\frac{D_{2n-1}(\al+2m{\textstyle\frac {1-\nu}{\nu}})}{D_{2(m+n)-1}(\al)}=\Gamma (\nu)
^{\frac {2m}\nu }(1-\nu)^{-m}\prod\limits _{j=1}^m\(\frac {m+n-j}{\frac \al 2 -j+\frac {2(n+m)-1}{2\nu}}\)\,,
\end{align*}
Further we have for $n>m$
\begin{align*}
&\frac{D_{2n-1}(2-\al-2m{\textstyle\frac {1-\nu}{\nu}})}{D_{2(n-m)-1}(2-\al)}=\Gamma (\nu)^{-\frac {2m}\nu }
(1-\nu)^{m}\prod\limits _{j=1}^m\(\frac {j-\frac \al 2 +\frac {2(n-m)-1}{2\nu}}{n-j}\)\,,
\end{align*}
and for $1<n\le m$
\begin{align*}
&D_{2n-1}(2-\al -2m{\textstyle\frac {1-\nu}{\nu}})D_{2(m-n)+1}(\al)=
\Gamma (\nu)^{-\frac {2m}\nu}(1-\nu)^m(-1)^{m-n+1}\\
&\times\frac i {\nu (n-1)!(m-n)!}\cot \frac\pi {2\nu}\(\nu \al+2(m-n)+1\)
\prod\limits _{j=1}^m \(j-\frac {\al} 2-\frac 1 {2\nu} (2(m-n)+1)\)\,.
\end{align*}

For $A=(a_1,\ldots,a_p),B=(b_1,\ldots,b_p)$ we set
\begin{align*}
\mathcal D(A|B|\kappa,\al)
=\det\left(\Theta\left(\textstyle{\frac{ia_j}{2\nu},\frac{ib_k}{2\nu}}|\kappa,\al\right)\right)_{j,k=1,\ldots,p}.
\end{align*}

Now, consider \eqref{idendesc}. Set
\begin{align*}
I^+=(a_1,\ldots,a_p),\ I^-=(b_1,\ldots,b_p),\ I^-_{<}=(b_1,\ldots,b_r),\ I^-_{>}=(b_{r+1},\ldots,b_p),
\end{align*}
where $a_1<\cdots<a_p$ and $b_1<\cdots<b_r<2m<b_{r+1}<\cdots<b_p$.
Using the above formulae and \eqref{par1}, \eqref{par3},  and \eqref{conjcomm},
one easily finds that \eqref{idendesc} is equivalent to the compatibility condition
\begin{align}
&F\cdot\mathcal{D}(I^+|I^-|\kappa, \al +2m{\textstyle\frac{1-\nu}\nu})
\mathcal{D}(\Io|\nu(\al-2I(m))|\kappa, \al )\label{eqdes2}\\
&=\mathcal{D}\left(\left(\Io\backslash(2m-I^-_<)\right)\sqcup (I^++2m)|
\nu(\al-2I(m))\sqcup\left(I^-_>-2m\right)|\kappa, \al\right),\nn
\end{align}
where
\begin{align*}
F&=(-2)^{-rm}(-\nu)^{-r}\prod_{j=1}^p\prod_{k=1}^m\frac{a_j+2m-(2k-1)}{a_j+2m+\nu(\al-2k)}
\prod_{j=r+1}^p\prod_{k=1}^m\frac{b_j-2m-\nu(\al-2k)}{b_j-2m+2k-1}\\
&\times\prod_{j=1}^r\prod_{k=1}^m(2m-b_j+\nu(\al-2k))\cdot
\prod_{j=1}^r\frac1{\left(\frac{b_j-1}2\right)!\left(m-\frac{b_j+1}2\right)!}.
%\frac i\nu\cot{\textstyle\frac\pi{2\nu}}(\nu\al+2m-b_j).
\end{align*}
It is already remarkable that $F$ is independent of $\kappa$ because it implies that the identity \eqref{eqdes2}
is valid for $\kappa=\infty$, where the function $\Theta(l,m|\kappa,\al)$ reduces to $-\frac i{l+m}$.
One can easily check the identity \eqref{eqdes2} for $\kappa=\infty$.
This implies that the compatibility condition is equivalent to
\begin{align}
&\frac{\mathcal{D}(I^+|I^-|\kappa, \al +2m{\textstyle\frac{1-\nu}\nu})}
{\mathcal{D}(I^+|I^-|\infty, \al +2m{\textstyle\frac{1-\nu}\nu})}\cdot
\frac{\mathcal{D}(\Io|\nu(\al-2I(m))|\kappa, \al )}{\mathcal{D}(\Io|\nu(\al-2I(m))|\infty, \al )}\label{eqdes}\\
&=\frac{\mathcal{D}\left(\left(\Io\backslash(2m-I^-_<)\right)\sqcup (I^++2m)|
\nu(\al-2I(m))\sqcup\left(I^-_>-2m\right)|\kappa, \al\right)}
{\mathcal{D}\left(\left(\Io\backslash(2m-I^-_<)\right)\sqcup (I^++2m)|
\nu(\al-2I(m))\sqcup\left(I^-_>-2m\right)|\infty, \al\right)}.\nn
\end{align}
We %check 
have checked
\eqref{eqdes} for $m=1$ up to level 8 and $\kappa ^{-8}$. 
We refer the reader to section \ref{cons}
for the proof of 
similar identities in the sG case.

\section{Gluing two chiralities}\label{two}

%The goal in this section is to compare the primary fields in the full CFT theory
%by using the actions of the fermionic operators. Let us consider the simplest case.
We have seen that, in the chiral CFT theory, the shifted primary field $\phi_{\al+2\frac{1-\nu}\nu}(0)$
is realised in $\mathcal H_\al$ as
\begin{align}
\phi_{\al+2\frac{1-\nu}\nu}(0)\cong\betab^*_1\gammab^*_{{\rm screen},1}\phi_\al(0).\label{def1}
\end{align}
The primary field in the left hand side is normalised by this formula.
We have a similar formula for the second chirality, %which we shall discuss later in more details.
\begin{align}
\bar\phi_{\al+2\frac{1-\nu}\nu}(0)\cong
\bar\gammab^*_1
\bar\betab^*_{{\rm screen},1}\bar\phi_\al(0).
\label{def2}
\end{align}
Consider the primary field $\Phi_\al(0)$ in the full CFT
normalised by the CFT theory. 
We have a relation
$$\Phi _{\al}(0)=S(\al)\phi _{\al}(0)\bar{\phi}_{\al}(0)\,,$$
for some function $S(\al)$. 
We shall determine
the exact ratio between 
$\Phi_{\al+2\frac{1-\nu}\nu}(0)$ and 
$\betab^*_1\gammab^*_{{\rm screen},1}
\bar\gammab^*_1\bar\betab^*_{{\rm screen},1}\Phi_\al(0)$. 
This, together with the definitions \eqref{def1}, 
\eqref{def2} and natural analyticity assumptions,  
allows one to determine $S(\al)$. 
However, we shall not write an explicit formula for $S(\al)$ 
because for the application  to the OPE 
what we need are the ratios (see \eqref{RATIO}).

The normalised three point function in full CFT can be extracted, for example, 
from the known results in Liouville theory. 
We quote from Appendix \ref{app1}, \eqref{THREEPOINT} and \eqref{xx}, which lead to
the following formula:
\begin{align}
&
\frac{\langle \Phi_{1-\kappa}(-\infty)\Phi_{\alpha+2\frac{1-\nu}{\nu}}(0)\Phi_{1+\kappa}(\infty)\rangle}
{\langle \Phi_{1-\kappa}(-\infty)\Phi_{\alpha}(0)\Phi_{1+\kappa}(\infty)\rangle}
=
\mub ^2 \Gamma (\nu)^2\cdot Y(x)\ W(\al,\kappa)\overline{W}(\al,\kappa)
\,,
\label{3pt}
\end{align}

where
\begin{align}
&x=\frac{\al}{2}+\frac{1-\nu}{2\nu}\,,
\label{def-x}
\end{align}
and we have set
\begin{align*}
&Y(x)=-2\nu x\cdot\frac {\Gamma ^2(\nu x +1/2-\nu/2)\Gamma (\nu-2\nu x)}
{\Gamma ^2(1/2+\nu/2-\nu x)\Gamma (2\nu x +1-\nu)}
\cdot\frac {\Gamma (-2\nu x)}{\Gamma (2\nu x)}\,,
\\
&W(\al,\kappa)=\frac{\Gamma(\al\nu/2-\nu+1+\kappa\nu)}{\Gamma(-\al\nu/2+\nu+\kappa\nu)}\,,\nn
\\
&\overline{W}(\al,\kappa)=W(\al,-\kappa)\,.\nn
\end{align*}

Now let us compare this result with the corresponding % correspomding
 one in the chiral theory.
Using \eqref{par3}, 
we immediately obtain:
\begin{align*}
&\frac{\langle 1-\kappa|\ \betab ^*_1\gammab ^*_{\mathrm{screen},1}
\phi _{\al}(0)\ |1+\kappa\rangle}
{\langle 1-\kappa|\ \phi _{\al}(0)\ |1+\kappa\rangle}=e^{\frac{\pi i}2
(2\nu-\al\nu -1)} %X(\al/2+(1-\nu)/2\nu)\\
X(x)
\\
&\times\(\nu\kappa\)^{\al\nu-2\nu+1}
 \Bigl(1-\frac{\al}{2}-\frac{1}{2\nu}\Bigr)
\Theta \Bigl(\frac{i}{2\nu}, -i(1-\frac{\al}{2})\Bigl|\kappa,\al \Bigr)\,,
\end{align*}
where
$$ X(x)=-i\Gamma (\nu)^{-2x+1}\cdot\frac {\Gamma (x+1/2 )}{ \Gamma(x)  }\cdot   
2^{-2\nu x+\nu}\frac{\Gamma(-\nu x+\nu/2)}{\Gamma(-\nu x+\nu/2+1/2)}\,. $$

The first thing to consider is the 
essential, $\kappa$-dependent, function $W(\al,\kappa)$. 
%We have for the asymptotics at $\kappa\to\infty$:
The asymptotics %for 
of
$W(\al,\kappa)$ for $\kappa\to\infty$ is given by
\begin{align}
W(\al,\kappa)
\simeq \(\kappa\nu\)^{\al\nu-2\nu+1}
\exp \Bigl(-\sum\limits_{k=1}^{\infty}\frac 1 {k(2k+1)}\frac 1 {(\kappa\nu)^{2k}}
B_{2k+1}(\al\nu/2 -\nu +1)\Bigr)\,.\nn
\end{align}
where $B_{n}(z)$ denotes the Bernoulli polynomial. 
On the other hand, using the asymptotic expansion
of $\Theta(l,m|\kappa,\alpha)$ available up to $\kappa ^{-8}$, we have checked that to this order
\begin{align}
&
 \Bigl(1-\frac{\al}{2}-\frac{1}{2\nu}\Bigr)
\Theta \Bigl(\frac{i}{2\nu}, -i(1-\frac{\al}{2})\Bigl|\kappa,\al \Bigr)
%(1-\al/2-1/2\nu)\Theta (i/2\nu, -i(1-\al/2)\ |\kappa,\al)
\label{comp}\\
&\simeq \exp \Bigl(-\sum\limits_{k=1}^{\infty}\frac 1 {k(2k+1)}\frac 1 {(\kappa\nu)^{2k}}
B_{2k+1}(\al\nu/2 -\nu +1)\Bigr)\,.\nn
\end{align}
So, we see that the $\kappa$-dependence agrees 
%there is an agreement 
between the chiral three point functions
computed in our way with the one  %that 
computed in CFT:
\begin{align}
&\frac{\langle 1-\kappa|\ \betab ^*_1\gammab ^*_{\mathrm{screen},1}
\phi _{\al}(0)\ |1+\kappa\rangle}{\langle 1-\kappa|\ \phi _{\al}(0)\ |1+\kappa\rangle}=e^{\frac{\pi i}2
(2\nu-\al\nu -1)}X(x)W(\al,\kappa)\,.
\label{1main}
\end{align}
%Here and after in the rest of this section, we use $x=\frac\al2+\frac{1-\nu}\nu$.

{Similarly we have for the second chirality that 
\begin{align}
&
\frac{\langle 1-\kappa|\ \bar\betab^*_{\mathrm{screen},1}\bar\gammab^*_{1}
\bar\phi_{\al}(0)\ |1+\kappa\rangle}
{\langle 1-\kappa|\ \bar\phi_{\al}(0)\ |1+\kappa\rangle}
=e^{-\frac{\pi i}2(2\nu-\al\nu -1)}X(x)\overline W(\al,\kappa)\,.
\label{2main}
\end{align}

Combining \eqref{3pt} with
\eqref{1main}, \eqref{2main}  %So, having in mind \eqref{comp} 
we obtain:
$$\Phi _{\al +2\frac{1-\nu}\nu}(0)\cong%=e^{Q-\bar{Q}}
C_1(\al)
%\pi \mub \gamma (\nu)\cdot C(\al/2+(1-\nu)/2\nu)
\betab ^*_1\bar{\gammab} ^*_{1}
\Phi _{\al}^{(1)}(0)\,,
$$
where
$$\Phi _{\al}^{(1)}(0)=i\mub^2\cot\textstyle\frac{\pi\nu}2
(2-\al )\bar{\betab }^*
_{\mathrm{screen},1}\gammab ^*_{\mathrm{screen},1}\Phi _{\al}(0)
$$
is one time screened primary field, and
$$
C_1(\al)= 
\Gamma (\nu)^2 Y(x)X(x)^{-2}i\cot \pi(\nu x -\frac{\nu}{2})\,,
%\cdot Y(\al/2+(1-\nu)/2\nu)X^{-2}(\al/2+(1-\nu)/2\nu)
%i\tan\textstyle\frac{\pi\nu}2(2-\al )\,.
$$
%We compute
%\begin{align}
%&C(x)=
%Y(x)X^{-2}(x)i\cot \pi(\nu x -\nu /2)
%=\nu \Gamma (\nu)^{4x-2} 2^{4\nu x-2\nu+1}\cdot\frac {\Gamma ^2(\nu x +1/2-\nu/2)\Gamma (\nu-2\nu x)}
%{\Gamma ^2(1/2+\nu/2-\nu x)\Gamma (2\nu x +1-\nu)}
%\nn\\ &\times 
%\frac
%   {\Gamma ^2( -\nu x+\nu/2  +1/2  )}{\Gamma ^2( -\nu x+\nu/2)    }\cdot  \frac{ x \Gamma ^2(x)  } {\Gamma ^2(x+1/2 )}\cdot
%\frac {\Gamma (-2\nu x)}
%{\Gamma (2\nu x)}i\cot \pi(\nu x -\nu /2)\,.\nn
%\end{align}
%Using  
there is a change of sign coming from different order
of fermions in the formulae \eqref{def2} and \eqref{2main}.

After simplification, 
we arrive at the result
%$$\Gamma ^2(-\nu x +\nu/2)\Gamma ^2(1/2-\nu x +\nu/2)= 2^{4\nu x -2\nu+2}\pi\ \Gamma ^2(-2\nu x+\nu)\,$$
%we simplify further
%\begin{align}
%%C(x)
%&Y(x)X^{-2}(x)i\cot \pi(\nu x -\nu /2)
%\label{YX-2}\\&=\nu \Gamma (\nu)^{4x-2}\ \frac {\Gamma (-2\nu x)}
%{\Gamma (2\nu x)}\cdot\frac{  \Gamma (x)  }{\Gamma (x+1/2 )}
%\cdot\frac{\Gamma (-x+1/2 )}{  \Gamma (-x)  }i\cot\pi x \,.
%
%\nn
%\end{align}
\begin{align}
C_1(\al)=-\nu \Gamma (\nu)^{4x}\ \frac {\Gamma (-2\nu x)}
{\Gamma (2\nu x)}\cdot\frac{  \Gamma (x)  }{\Gamma (x+1/2 )}
\cdot\frac{\Gamma (-x+1/2 )}{  \Gamma (-x)  }i\cot\pi x \,,
\label{C1}
\end{align}
where $x$ is defined in \eqref{def-x}. 
Being just a normalisation of primary fields,  
this formula may not look very significant.
%This formula may look not very significant 
%being just a normalisation of primary fields. 
%However, the main advantage of our approach is
%that it allows simple computation of one-point functions
%for the sG model. 
%$We 
Nevertheless, we
shall see that the formula 
\eqref{C1}
%\eqref{YX-2}
implies the
Lukyanov-Zamolodchikov
formula for the one-point functions of primary fields \cite{LukZam}
(see subsection \ref{LZformula}).
We remark that
\begin{align}
\frac{S(\al+2\frac{1-\nu}\nu)}{S(\al)}=i\mub^2
\cot {\textstyle\frac {\pi\nu}2}(2-\al)
C_1(\al).\label{RATIO}
\end{align}

It is easy to generalise the calculation given above. For any $m>0$ we have
\begin{align}
&\Phi _{\al +2m\frac{1-\nu}\nu}(0)\cong%=e^{m(Q-\bar{Q})}
C_m(\al)\betab ^*_{\Io}\bar{\gammab }^*_{\Io}
%\gammab ^*_{\mathrm{screen},\{m\}}\bar{\betab} ^*_{\mathrm{screen}, \{m\}}
\Phi _{\al}^{(m)}(0)\,,\label{TWOPRI}
\end{align}
where
\begin{align}
&\Phi _{\al}^{(m)}(0)=i^m\mub ^{2m}\prod\limits _{j=1}^{m}\cot {\textstyle\frac {\pi\nu}2}(2j-\al)
\bar{\betab}^*_{\mathrm{screen},I(m)}\gammab^*_{\mathrm{screen},I(m)}\Phi _{\al}(0)\,,\label{calpha}\\
&C_m(\al)=\prod\limits_{j=0}^{m-1}C_1(\al+2j
{\textstyle \frac {1-\nu}\nu})\,.
\nn
\end{align}
The multiplier is included in order that
the one-point function of $\Phi _{\al}^{(m)}(0)$ be a simple power in $\mub$ in the sG case (see \eqref{NORMALIZATION}).
The formula \eqref{TWOPRI} generalises \eqref{PRIMARY} to the case of two chiralities.
The entire space $
\mathcal{V}^\mathrm{quo}_{\al +2m\frac {1-\nu}\nu}\otimes
\bar{\mathcal{V}}^\mathrm{quo}_{\al +2m\frac {1-\nu}\nu}$ is embedded 
into the space $\mathcal{H}_{\al}\otimes \bar{\mathcal{H}}_{\al}$ the basis being
$$\betab^* _{I^+}\bar{\betab} ^*_{\bar{I}^+}
\bar{\gammab }^*_{\bar{I}^-}\gammab^*_{I^-}
%\ \gammab ^*_{\mathrm{screen}, \{m\}}%\cdots\gammab ^*_{\mathrm{screen}, m}
%\ \bar{\betab}^*_{\mathrm{screen},\{m\}}
\Phi^{(m)}_{\al}(0)\,.$$
The identification of descendants (see \eqref{idendesc}) in the case of two chiralities reads as
\begin{align}
&\betab^*_{I^+}\bar\betab^*_{\bar{I}^+}
\bar\gammab^*_{\bar{I}^-}\gammab^*_{I^-}
\Phi_{\alpha+2m\frac{1-\nu}{\nu}}(0)\label{IDENDESC}\\
 &\cong C_m(\al)\betab^*_{I^+ +2m}\bar\betab^*_{\bar{I}^+-2m}
\bar\gammab^*_{\bar{I}^- +2m}\gammab^*_{I^--2m}\betab^*_{\Io}\bar\gammab^*_{\Io}
\Phi^{(m)}_{\alpha}(0).\nn
\end{align}

%%%%%%%%%%%%%%%%%%%%%%%%%%%%%%%

\section{ Creation operators in sG case and function $\omega _R^\mathrm{sG}(\z,\xi|\al)$}
\label{TBA}

Before embarking upon the scaling limit to the sG model, 
let us give a brief review about the creation operators in the inhomogeneous case. For the inhomogeneous model 
the annihilation operators 
split into two parts:
\begin{align}
&\bb (\z)=\bb ^+(\z)+\bb ^-(\z), \qquad \cb (\z)=\cb ^+(\z)+\cb ^-(\z)\,,
\nn\\
&\bb ^{\pm}(\z)=\sum\limits _{p=1}^\infty %\(\z^2\z_0^{\mp 2}-1\)^{-p+1}
\(\z^2\z_0^{\mp 2}-1\)^{-p}
\bb^\pm_p,
\qquad \cb ^{\pm}(\z)=\sum\limits _{p=1}^\infty %\(\z^2\z_0^{\mp 2}-1\)^{-p+1}
\(\z^2\z_0^{\mp 2}-1\)^{-p}
\cb^\pm_p\,.
\nn
\end{align}
Consider the creation operator $\bb ^*_0(\z)$. The local operators are created
by its power series at $\z ^2=\z_0^{\pm 2}$:
\begin{align}
\bb ^{*}_0(\z)
\ \ \simeq\hskip -.8cm{}_{{}_{{}_{\scalebox{.7}
{$\z^2\to \z _0^{\pm2}$}}}}
\ \ 
\sum_{p=1}^{\infty}
(\z ^2\z _0^{\mp2}-1)^{p-1}\bb ^{\pm *}_{0,p}\,,
\nn
\end{align}
we denote the corresponding sums by
\begin{align}
\bb ^{\pm *}_0(\z)=\sum_{p=1}^{\infty}
(\z ^2\z _0^{\mp2}-1)^{p-1}\bb ^{\pm *}_{0,p}\,.
\nn
\end{align}
Then we introduce the notation
\begin{align}
\Omega _0^{\epsilon\epsilon'}=
\frac{4}{(2\pi i)^2}
\int\limits _{\Gamma_{\epsilon}}
\int\limits _{\Gamma   _{\epsilon'}  }\omega (\z/\xi,\al)
\bb^{\epsilon} (\z)\cb ^{\epsilon'}(\xi)\frac{d\z^2}{\z ^2}\ \frac{d\xi^2}{\xi ^2}\,,
\nn
\end{align}
and define
\begin{align}
&\bb ^{+*}(\z)
=e^{-\Omega _)^{++}}\bb ^{+*}_0(\z)e^{\Omega _0^{++}}
\,,
\quad \cb ^{+*}(\z)
=e^{-\Omega _0^{++}}\cb ^{+*}_0(\z)e^{\Omega _0^{++}}\,,
\label{defbcsG}
\\
&\bb ^{-*}(\z)
=e^{-\Omega _0^{--}}\bb ^{-*}_0(\z)e^{\Omega _0^{--}}
\,,
\quad \cb ^{-*}(\z)
=e^{-\Omega _0^{--}}\cb ^{-*}_0(\z)e^{\Omega _0^{--}}\,.
\nn
\end{align}
The functional $Z^{-s}_\mathbf{n}$ computed on the descendants created by 
these operators takes the determinant form \cite{HGSII,HGSIII} with the pairing
being:
\begin{align}
&Z^{-s}_\mathbf{n}\{\bb ^{+*}(\z)\cb ^{+*}(\xi)
\bigl(q^{2\al S(0)}\bigr)\}=\omega _\mathbf{n}(\z,\xi|\al,
-s)\,,\label{latpar}\\
&Z^{-s}_\mathbf{n}\{\bb ^{+*}(\z)\cb ^{-*}(\xi)
\bigl(q^{2\al S(0)}\bigr)\}=\omega _\mathbf{n}(\z,\xi|\al,-s)
+\omega_0 (\z/\xi,\al)%+\omega_0 (\z/\xi|\al)
\,,\nn\\
&Z^{-s}_\mathbf{n}\{\bb ^{-*}(\z)\cb ^{+*}(\xi)
\bigl(q^{2\al S(0)}\bigr)\}=\omega _\mathbf{n}(\z,\xi|\al,-s)
+\omega_0 (\z/\xi,\al)%+\omega_0 (\z/\xi|\al)
\,,\nn\\
&Z^{-s}_\mathbf{n}\{\bb ^{-*}(\z)\cb ^{-*}(\xi)
\bigl(q^{2\al S(0)}\bigr)\}=\omega _\mathbf{n}(\z,\xi|\al,-s)
\,.\nn
\end{align}
We do not give the definition of the function $\omega _\mathbf{n}(\z,\xi|\al,-s)$
which is easy to find from \cite{HGSIV} because we shall be interested only in its scaling limit. 

We conjecture that the following scaling limit 
exists for the operators $\bb ^{\pm*}(\z)$, $\cb ^{\pm*}(\z)$:
\begin{align*}
&\half\bb ^{+*}(\z)\ \ \longrightarrow
\hskip -1cm {}_{{\ }_{\mathrm{scaling}}}
\ \ \betab^{+*}(\z)\simeq
\betab^*(\mub\z)+\bar{\betab}^*_\mathrm{screen}
(\z/\mub),\quad \z \to\infty\,,
\\&
\half\cb ^{+*}(\z)   \ \ \longrightarrow
\hskip -1cm {}_{{\ }_{\mathrm{scaling}}}
\ \ \gammab^{+*}(\z)\simeq
\gammab^*(\mub\z)+\bar{\gammab}^*_\mathrm{screen}
(\z/\mub),\quad\  \z \to\infty\,,\\
&\half\bb ^{-*}(\z)\ \ \longrightarrow
\hskip -1cm {}_{{\ }_{\mathrm{scaling}}}
\ \ \betab^{-*}(\z)\simeq\bar{\betab}^*(\z/\mub)+{\betab}^*_\mathrm{screen}
(\mub \z),\ \quad \z \to 0\,,\\
&\half\cb ^{-*}(\z)\ \ \longrightarrow
\hskip -1cm {}_{{\ }_{\mathrm{scaling}}}
\ \ \gammab^{-*}(\z)\simeq\bar{\gammab}^*(\z/\mub)+\gammab^*_\mathrm{screen}
(\mub \z),\ \ \quad \z \to 0\,.
\end{align*}
These operators have the asymptotics \eqref{betgam}, \eqref{scr1},
\eqref{BAREXPANSION1}, \eqref{BAREXPANSION2}.
An explanation to this conjecture is given in \cite{OP}. Notice that the appearance of $\mub$ in the right
hand sides in not a part of the conjecture, but is a corollary of the computations done in the
conformal case \cite{HGSIV}: the key identity is
\begin{align*}
\z _0(Ca)^\nu=\mub^{-1}.
\end{align*}

%%%%%%%%%%%%%%
In this section we define our main function $\omega _R^\mathrm{sG}(\z,\xi|\al)$ as the scaling limit of $\omega _\mathbf{n}(\z,\xi|\al,s)$
 for $\al=2s{\textstyle\frac {1-\nu}{\nu}}$, 
and as the analytic continuation with respect to $\alpha$ in general. We shall be rather sketchy because 
the construction repeats very much what has been done in the conformal case \cite{HGSIV}.

We start with  the DDV equation:
\begin{align}
&\frac 1 i \log \mathfrak{a}(\z)=
\pi MR (\z ^{1/\nu}-\z ^{-1/\nu})
%\label{DDV}\\&+
%i 
-2\mathrm{Im}\int\limits _{0}^{\infty }
%e^{i0}
R (\z/\xi) \log
(1+\mathfrak{a}(\xi e^{+i 0}))
%\log(1+\mathfrak{a}(\xi))
\frac {d\xi ^2}{\xi ^2}
%-i \int\limits _{0}^{\infty e^{-i0}}R (\z/\xi) \log
%(1+\bar{\mathfrak{a}}(\xi))\frac {d\xi ^2}{\xi ^2}
\,,\label{DDV}
\end{align}
where as usual it is convenient to define $R(\z)$ through a more general object:
\begin{align}
&R (\z,\al)=\int\limits _{-\infty}^{\infty} \z ^{2ik}\widehat{R}(k,\al)
\frac{dk}{2\pi},\quad\widehat{R}(k,\al)=
\frac {\sinh\pi((2\nu-1)k-i\al/2)}{2\sinh\pi ((1-\nu)k+i\al/2)\cosh (\pi\nu k)}\,,\nn\\
&R (\z)=R (\z,0)\,.\nn
\end{align}
Notice that
\begin{align}
\widehat{R}(k,-\al)=\widehat{R}(-k,\al),\quad \widehat{R}(k,\al+2)=\widehat{R}(k,\al)\,.
\label{R=R}
\end{align}

We hope using the same letter for the resolvent as in the conformal case \cite{HGSIV}
is not very confusing. Similarly to \cite{HGSIV}, we write the equation for the resolvent:
\begin{align}
R_{\mathrm{dress}}+R\ast R_{\mathrm{dress}}=R\,,\label{eqres}
\end{align}
and define $\omega ^{\mathrm{sG}}_R$ by
\begin{align}
\frac{1}{2\pi i} \ \omega^{\mathrm{sG}}_R =-F^+\ast F^- +F^+\ast R_{\mathrm{dress}}\ast F^-\,,
\label{eqomega}
\end{align}
where
$$f\ast g=\int_{0}^{\infty}f(\z)g(\z)dm(\z),\quad 
dm(\z)=2\mathrm{Re}\(\frac 1 {1+\mathfrak{a}(\z e^{-i0})}\)\frac {d\z ^2}
{\z^2}\,,$$
and
$$F^\pm(\z,\xi)=F^\pm(\z/\xi),
\quad F^\pm(\z)=\frac {i}{
2\pi  \nu(\z ^{\frac 1 {\nu}}e^{\mp i0}-\z^{-\frac 1 {\nu}})}=\mp \int\frac{dk}{2\pi}
\z^{2ik}\frac{e^{\mp \pi \nu k}}{2\cosh(\pi \nu k)}
%=i\frac {\la ^{1/\nu}}{
%\nu(1+\la ^{2/\nu})}, \quad \widehat{F}(k)=\frac {2\pi i} {2\cosh(\pi\nu k)}
\,.$$
It is very convenient to consider not the function 
$\omega^{\mathrm{sG}}_R(\z,\xi|\al)$, but rather
its Mellin transform. Namely, we introduce $\Theta ^{\mathrm{sG}}_R(l,m|\al)$ by
\begin{align}
R_{\mathrm{dress}}(\z,\xi)
-R(\z/\xi,\al)=\int\limits_{-\infty}^{\infty}
\int\limits_{-\infty}^{\infty}\frac{dl}{2\pi}\frac{dm}{2\pi} \ 
\widehat{R}(l,\al)\Theta ^{\mathrm{sG}}_R(l,m|\al)\widehat{R}(m,-\al)
\z ^{2il}\xi ^{2im}\,.\nn
\end{align}
%Using  the simple-minded formula
%\begin{align}
%R(\z/\xi,\al)=\int\limits_{-\infty}^{\infty}
%\int\limits_{-\infty}^{\infty}\frac{dl}{2\pi}\frac{dm}{2\pi} \ 
%\widehat{R}(l,\al)\frac{2\pi \delta (l+m)}{\widehat{R}(m,-\al)}
%\widehat{R}(m,-\al)
%\z ^{2il}\xi ^{2im}\,,\nn
%\end{align}
Rewriting \eqref{eqres}
we get the equation for $\Theta^{\mathrm{sG}}_R$:
\begin{align}
\Theta ^{\mathrm{sG}}_R(l,m|\al)+G(l+m)+\int\limits _{-\infty}^{\infty}G(l-k)\widehat{R}
(k,\al)\Theta ^{\mathrm{sG}}_R(k,m|\al)\frac{dk}{2\pi } =0\,,\label{eqTheta}
\end{align}
where $G(k)$ is the moment of our measure:
$$G(k)=\int_{0}^{\infty}\z ^{-2ik}dm(\z)\,.$$
There
are no obstacles for the convergence of the integral
in the entire complex plane of $k$, so, $G(k)$ is an entire function.

Now it is rather easy to see that
\begin{align}
\omega ^{\mathrm{sG}}_R(\z,\xi|\al)=%2\pi i
-\frac{\pi i}{2}
\int\limits _{-\infty}^{\infty}\int\limits _{-\infty}^{\infty}
\frac{dl}{2\pi}\frac{dm}{2\pi} 
\z ^{2il}\xi ^{2im}
\frac {e^{-\pi \nu l}}{\cosh(\pi\nu l)}
\Theta ^{\mathrm{sG}}_R(l,m|\al)
\frac {e^{-\pi \nu m}} {\cosh(\pi\nu m)}
\,.
\label{finomega}
\end{align}
What we are really interested in are
the coefficients in the asymptotic expansion
\begin{align}
\omega^{\mathrm{sG}} _R(\z,\xi|\al)
%\simeq_{\la\to\epsilon_1\infty,\xi\to\epsilon _2\infty}
\simeq\sum\limits _{j,k=1}^{\infty}
\z ^{-\epsilon _1\frac{2j-1}{\nu}}\xi^{-\epsilon _2\frac{2k-1}{\nu}}
\omega^{\mathrm{sG}} _R{}_{\epsilon_1(2j-1),\epsilon _2(2k-1)}(\al)%\,.
\qquad
(\z^{\epsilon_1},\xi^{\epsilon_2}\to\infty)\,,
\end{align}
where $\epsilon _1,\epsilon _2=\pm 1$.
Obviously,
$$\omega^{\mathrm{sG}}_R{} _{2j-1,2k-1}(\al)=\mathrm{sgn}(2j-1)
\mathrm{sgn}(2k-1)
\frac {i}{2\pi  \nu^2}
\Theta ^{\mathrm{sG}}_R\Bigl({\textstyle\frac{2j-1}{2\nu}}i,
{\textstyle \frac{2k-1}{2\nu}}i|
\al\Bigr)\,,$$
for all odd integer  $2j-1$, $2k-1$.

The function $\mathfrak{a}(\z)$ possesses the symmetry
\begin{align}
&\mathfrak{a}(\z)=\(\mathfrak{a}(\z^{-1})\)^{-1}\,.\label{syma}
\end{align}
This property implies that $G(k)$ is an even function.
Together with
%Then using the following consequence of \eqref{R=R}
$$\hat{R}(k,2-\al)=\hat{R}(-k,\al)\,,$$
which follows from \eqref{R=R}, 
%we prove 
we then obtain
the symmetry
\begin{align}
\Theta ^{\mathrm{sG}}_R(l,m|2-\al)=\Theta ^{\mathrm{sG}}_R(-l,-m|\al)\,.\label{symth}
\end{align}
The position of the singularities is important in the derivation.
This symmetry property provides, for example,  the invariance  of our main formula \eqref{themain}
derived in the next section under the interchange of 
the two chiralities.

Notice that the symmetry  \eqref{syma} is 
a property specific to the maximal eigenvalue 
of the Matsubara transfer-matrix. 
Certainly, this is the only case interesting for us, but 
most of our computations hold true if we consider 
$\mathfrak{a}(\z)$ corresponding to other eigenvalues
as well. In the
latter case the symmetry \eqref{syma} is broken. Since
considering other eigenvalues may be of physical interest,
in the following calculations we shall not  use the fact that $G(k)$ is even.

\section{Main formula}\label{sec:main}

Following  \cite{OP} we conclude that the one-point functions
at finite $R$ are described by usual fermionic formulae.
For the same chiralities we have to take $\omega^{\mathrm{sG}}_R(\z,\xi|\al)$ while for different
chiralities we have to take $\omega^{\mathrm{sG}}_R(\z,\xi|\al)+\omega_0(\z/\xi,\al)$:
\begin{align}
&\frac {\langle \betab ^{+*}(\z)\gammab^{+*}(\xi) \Phi _\al (0)\rangle ^\mathrm{sG}_R}
{\langle  \Phi _\al (0)\rangle ^\mathrm{sG}_R}=\omega^\mathrm{sG}_R
(\z,\xi|\al)\,,\nn\\
&\frac {\langle \betab ^{+*}(\z)\gammab^{-*}(\xi) \Phi _\al (0)\rangle ^\mathrm{sG}_R}
{\langle  \Phi _\al (0)\rangle ^\mathrm{sG}_R}=\omega^\mathrm{sG}_R
(\z,\xi|\al)+\omega_0(\z/\xi,\al)\,,\nn\\
&\frac {\langle \betab ^{-*}(\z)\gammab^{+*}(\xi) \Phi _\al (0)\rangle ^\mathrm{sG}_R}
{\langle  \Phi _\al (0)\rangle ^\mathrm{sG}_R}=\omega^\mathrm{sG}_R
(\z,\xi|\al)+\omega_0(\z/\xi,\al)\,,\nn\\
&\frac {\langle \betab ^{-*}(\z)\gammab^{-*}(\xi) \Phi _\al (0)\rangle ^\mathrm{sG}_R}
{\langle  \Phi _\al (0)\rangle ^\mathrm{sG}_R}=\omega^\mathrm{sG}_R
(\z,\xi|\al)\,.\nn
\end{align}
We have computed the asymptotics of $\omega^\mathrm{sG}_R(\z,\xi|\al)$
in the previous section. Now we compute the asymptotics of $\omega_0(\z,\al)$:
\begin{align}
&\omega_0 (\z,\al)=- i\int\limits _{-\infty }^{\infty }\z ^{2ik}\frac{\sinh\frac{\pi}2(2(1-\nu)k+i\al)}
{2\sinh\frac{\pi}2(2k+i\al)\cosh\pi\nu k }dk\label{asymomega}\\
&\simeq\frac {i\epsilon}{\nu}\sum\limits _{j=1}^{\infty}
\z ^{-\frac{2j-1}\nu\epsilon}\cot \frac {\pi}{2\nu}(\nu \al +(2j-1)
\epsilon) +i\epsilon\sum\limits _{j=1}^{\infty}\z^{\al-1-(2j-1)\epsilon}\tan \frac {\pi \nu}2(\al-1-(2j-1)\epsilon),\nn\\
&\hskip300pt\text{for $\z^\epsilon\rightarrow\infty$.}\nn
\end{align}

From the asymptotics one can read the expectation values of the fermion operators.
There is a difference between the CFT case and the sG case. Two cases have different selection rules.
In the CFT case, the expectation values are zero between chiral and anti-chiral components,
while there are non-zero values between screening and non-screening operators. This is opposite in the
sG case. Let us list some of the non-zero expectation values in the sG case. It is convenient to introduce
the convention: for $j\geq1$
\begin{align*}
\tilde\betab^*_{2j-1}=\betab^*_{2j-1},\ \tilde\betab^*_{1-2j}=\bar\betab^*_{2j-1},
\ \tilde\gammab^*_{2j-1}=\gammab^*_{2j-1},\ \tilde\gammab^*_{1-2j}=\bar\gammab^*_{2j-1}.
\end{align*}
Then, for $a,b\in2\mathbb{Z}+1$, we have
\begin{align*}
&\frac{\langle\tilde\betab^*_a\tilde\gammab^*_b\Phi_\al(0)\rangle^{\rm sG}_R}{\langle\Phi_\al(0)\rangle^{\rm sG}_R}\\
&\quad=\mathrm{sgn}(a)\,\mathrm{sgn}(b)\frac{i\mub^{\frac{a+b}\nu}}{2\pi\nu^2}
\left( \Theta ^{\mathrm{sG}}_R\({\textstyle \frac {ia}{2\nu}},{\textstyle \frac {ib}{2\nu}}|\al\)
-\mathrm{sgn}(a)\delta _{a,-b}2\pi\nu\cot
{\textstyle \frac {\pi}{2\nu }}(\nu\al+a)\right),
\end{align*}
and for $j,k\geq1$
\begin{align*}
\frac{\langle\bar\betab^*_{\mathrm{screen},j}\gammab^*_{\mathrm{screen},k}\Phi_\al(0)\rangle^{\rm sG}_R}
{\langle\Phi_\al(0)\rangle^{\rm sG}_R}
=i\mub^{\al-2j}\delta_{j,k}\tan{\textstyle\frac{\pi\nu}2}(\al-2j).
\end{align*}

Similarly to the conformal case let us introduce for $\#(A)=\#(B)=n$:
\begin{align*}
&\mathcal{D}^{\mathrm{sG}}_R(A|B|\al)=\prod\limits _{j=1}^n
\mathrm{sgn}(a_j)\mathrm{sgn}(b_j)\(\frac i{2\pi\nu^2}\)^n\det\( D_{a_n,b_k}(\al)\)|_{j,k=1,\cdots ,n},\\
&D_{a,b}(\al)=\Theta^{\mathrm sG}_R\bigl({\textstyle\frac{ia}{2\nu}},{\textstyle\frac{ib}{2\nu}}|\alpha\bigr)
-\delta_{a,-b}\mathrm{sgn}(a)2\pi\nu \cot{\textstyle\frac{\pi}{2\nu}}(\nu\alpha+a)\,,
\end{align*}
We have seen that the primary fields $\Phi _{\al +2m\frac {1-\nu}\nu}(0)$ and all their descendants
are obtained from
$$\betab^* _{I^+}\bar{\betab} ^*_{\bar{I}^+}
\bar{\gammab }^*_{\bar{I}^-}\gammab^*_{I^-}
\ \Phi^{(m)}_{\al}(0)\,.$$
So is is sufficient to write down the formula for one-point functions of these operators in the sG
model. This is immediate. The only important thing to notice is
that the contribution from screening operators completely
disappears as a result of our normalisation
of $\Phi ^{(m)}(0)$. 
Thus we have the main formula:
\begin{align}
&\frac {\langle
\betab^* _{I^+}\bar{\betab} ^*_{\bar{I}^+}
\bar{\gammab }^*_{\bar{I}^-}\gammab^*_{I^-}
%\ \gammab ^*_{\mathrm{screen}, \{m\}}%\cdots\gammab ^*_{\mathrm{screen}, m}
%\ \bar{\betab}^*_{\mathrm{screen},\{m\}}
\Phi ^{(m)}_{\al}(0)
 \rangle^\mathrm{sG}_R}
{\langle\Phi_{\al }(0) \rangle^\mathrm{sG}_R}\label{themain}\\
&=%(-1)^{\#(I^+)\cdot  \#(\bar{I}^-)}
%\left[ M\frac {\sqrt{\pi} \ \Gamma (\frac 1 {2\nu})}
%{2\Gamma (\frac {1-\nu}{2\nu})  }  \Gamma (\nu)^{-\frac 1 \nu}\right]^{2m\nu\al -2m^2\nu   +|I^+|+|I^-|+|\bar{I}^+|+|\bar{I}^-|}\nn
\mub ^{  2m\al -2m^2+\frac 1 {\nu}\(   |I^+|+|I^-|+|\bar{I}^+|+|\bar{I}^-|    \)      }
%+2\Delta _{\al+2\frac{1-\nu}\nu}-2\Delta_\al 
%\(\Gamma (\nu)\)^{-\frac 1 \nu( |I^+|+|I^-|+|\bar{I}^+|+|\bar{I}^-|)}\nn
%\nn\\&\times
%\prod\limits _{j=1}^{m}\tan {\textstyle \frac{\pi\nu}2(2j-\al)}
%\\ &\times
\ \mathcal{D}^{\mathrm{sG}}_R\(I^+\cup (-\bar{I}^+)\ |\ I^-\cup
(-\bar{I}^-)|\al\)\,,\nn
\end{align}
with the requirements $\#(I^+)=\#(I^-)+m$, $\#(\bar{I}^+)+m=\#(\bar{I}^-)$.
In particular, we have
\begin{align}
&\frac{\langle\Phi ^{(m)}_{\al}(0) \rangle^\mathrm{sG}_R}
{\langle\Phi_{\al }(0)\rangle^\mathrm{sG}_R}
=\mub^{2m\al-2m^2}.\label{NORMALIZATION}
\end{align}

\section{Proof of compatibility}\label{cons}

The main formula \eqref{themain} leads to the evaluation of
the expectation values for the primary fields and the descendants. Namely,
from \eqref{TWOPRI} and \eqref{IDENDESC}, we obtain
\begin{align}
&\frac{\langle \Phi_{\alpha+2m
\frac{1-\nu}{\nu}}(0)\rangle^\mathrm{sG}_R}{\langle \Phi_{\alpha}(0)\rangle^\mathrm{sG}_R}
=C_m(\alpha)\frac{\langle \betab^*_{\Io}\bar\gammab^*_{\Io}\Phi^{(m)}_{\alpha}(0)\rangle^\mathrm{sG}_R}
{\langle \Phi_{\alpha}(0)\rangle^\mathrm{sG}_R}\,,\label{eqtoprove}\\
&\frac{\langle \betab^*_{I^+}\bar\betab^*_{\bar{I}^+}\bar\gammab^*_{\bar{I}^-}\gammab^*_{I^-}
\Phi_{\alpha+2m\frac{1-\nu}{\nu}}(0)\rangle^\mathrm{sG}_R}
{\langle \Phi_{\alpha+2m\frac{1-\nu}{\nu}}(0)\rangle^\mathrm{sG}_R}\nn\\
&=\frac{\langle \betab^*_{I^+ +2m}\bar\betab^*_{\bar{I}^+-2m}
\bar\gammab^*_{\bar{I}^- +2m}\gammab^*_{I^--2m}\betab^*_{\Io}\bar\gammab^*_{\Io}
\Phi^{(m)}_{\alpha}(0)\rangle^\mathrm{sG}_R}
{\langle \betab^*_{\Io}\bar\gammab^*_{\Io}\Phi^{(m)}_{\alpha}(0)\rangle^\mathrm{sG}_R}\,,\nn
\end{align}
where $\#(I^+)=\#(I^-)$ and $\#(\bar I^+)=\#(\bar I^-)$.
In order for these equalities to hold,
certain consistency conditions need to be satisfied. 
Recalling the definition \eqref{PAIRING}, \eqref{TRANS}, \eqref{conjcomm}, we find that 
the conditions read respectively as follows. \begin{align}
&\det\left(D_{a,b}(0,\al)\right)_{a\in \Io, b\in -\Io
}=\prod_{j=0}^{m-1}D_{1,-1}(j,\al),
\label{1st-id}\\
&\det\left(D_{a,b}(0,\al)\right)_{a\in J^+, b\in J^-
%I_0(m)
}=(-1)^{\#}\prod\limits _{a\in (2m-I^-)_>\sqcup(2m-\bar{I}^+)_>}
2\pi\nu\cot{\textstyle\frac{\pi}{2\nu}}(\nu\alpha+a)
\label{2nd-id}\\&\qquad\quad
\times\det\left(D_{a,b}(m,\al)\right)_{a\in I^+\cup( -\bar{I}^+)
\atop b\in I^- \cup (-\bar{I}^-)}\ \prod_{j=0}^{m-1}D_{1,-1}(j,\al)\,,
\nn
\end{align}
where 
\begin{align}
&D_{a,b}(j,\al)=D_{a,b}(\al+2j{\textstyle\frac{1-\nu}\nu}),\nn\\
&J^+=(I^++2m)\cup (-\bar{I}^++2m)_<\cup (\Io \backslash (2m-I^-)_>) \,,\nn\\
&J^-= (I^--2m)_>\cup -(\bar{I}^- + 2m)\cup(-\Io\backslash (\bar{I}^+-2m)_<)\,,\nn
\end{align}
and the power of $-1$ can be easily computed from the fermionic
commutation relations. 
These identities are understood as analytic continuation from the region 
\begin{align*}
\alpha+2m{\textstyle \frac{1-\nu}{\nu}}<2.
\end{align*}

We show below that they are consequences of the single identity
\begin{align}
&\Theta^{\mathrm sG}_R(l,j|\alpha+2{\textstyle \frac{1-\nu}{\nu}})
-\Theta^{\mathrm sG}_R(l+{\textstyle\frac{i}{\nu}},j-{\textstyle\frac{i}{\nu}}|\alpha)
=-\frac{\Theta^{\mathrm sG}_R(l+\frac{i}{\nu},-\frac{i}{2\nu}|\alpha)
\Theta^{\mathrm sG}_R(\frac{i}{2\nu},j-\frac{i}{\nu}|\alpha)}
{\Theta^{\mathrm sG}_R(\frac{i}{2\nu},-\frac{i}{2\nu}|\alpha)
-2\pi\nu\cot\frac{\pi}{2}(\alpha+\frac{1}{\nu})
}\,,
\label{id-theta}%\nn
\end{align}
where we start from the case when both $\al$ and $\al+2
\frac {1-\nu}\nu$ 
are inside the interval $(0,2)$ and then continue analytically. 
First let us verify \eqref{id-theta}.  
We start from the defining equation \eqref{eqTheta} with shifted $\alpha$, 
\begin{align*}
&\Theta^{\mathrm sG}_R(l,j|\alpha+2{\textstyle \frac{1-\nu}{\nu}})
+G(l+j)
\\
&\quad+\int_{-\infty}^\infty G(l-k)\hat{R}(k,\alpha+2{\textstyle \frac{1-\nu}{\nu}})
\Theta^{\mathrm sG}_R(k,j|\alpha+2{\textstyle \frac{1-\nu}{\nu}})
\frac{dk}{2\pi}=0\,.
\end{align*}
Noting the relation
\begin{align*}
\hat{R}(k,\alpha+2{\textstyle \frac{1-\nu}{\nu}})=\hat{R}(k+
{\textstyle\frac{i}{\nu}},\alpha),
\end{align*}   
we shift the contour to $\mathrm{Im}\,k=-1/\nu$. 
Under our assumption, the only pole encountered on the way is  $k=-i/2\nu$. 
Denote by $X(l,j)$ the left hand side of \eqref{id-theta}.
Combining the above calculation with \eqref{eqTheta} we find 
\begin{align*}
&X(l,j)+\int_{-\infty}^\infty G(l-k+
{\textstyle\frac{i}{\nu}})\hat{R}(k,\alpha)X(k-{\textstyle\frac i\nu},j)\frac{dk}{2\pi}
\\
&\quad
+{\textstyle\frac{1}{2\pi\nu}}\tan
{\textstyle\frac{\pi}{2}}(\alpha+{\textstyle\frac{1}{\nu}})
G(l+
{\textstyle\frac{i}{2\nu}})\Theta^{\mathrm sG}_R(-
{\textstyle\frac{i}{2\nu}},j|\alpha+2{\textstyle \frac{1-\nu}{\nu}})=0\,,
\end{align*}
which can be solved as
\begin{align*}
X(l,j)
={\textstyle\frac{1}{2\pi\nu}}\tan
{\textstyle\frac{\pi}{2}}(\alpha+{\textstyle\frac{1}{\nu}})
\Theta^{\mathrm sG}_R(l+
{\textstyle\frac{i}{\nu}},-
{\textstyle\frac{i}{2\nu}}|\alpha)
\Theta^{\mathrm sG}_R(-
{\textstyle\frac{i}{2\nu}},j|\alpha+2{\textstyle \frac{1-\nu}{\nu}})\,.
\end{align*}
Setting $l=-i/2\nu$ in the last formula
and eliminating $\Theta^{\mathrm sG}_R(-
{\textstyle\frac{i}{2\nu}},j|\alpha+2{\textstyle \frac{1-\nu}{\nu}})$, 
we arrive at \eqref{id-theta}. 

Now we return to \eqref{1st-id}, \eqref{2nd-id}.  
Specialising \eqref{id-theta} to $l=ia/2\nu$ and $j=ib/2\nu$, we obtain
\begin{align}
&D_{a,b}(1,\al)D_{1,-1}(0,\al)=
\det
\begin{pmatrix}
D_{1,-1}(0,\al) & D_{1,b-2}(0,\al)\\
D_{a+2,-1}(0,\al) & D_{a+2,b-2}(0,\al) \\
\end{pmatrix},\quad\   a\ne -1,\ b\ne 1
\,,\label{basic-id}\\
&D_{-1,b}(1,\al)D_{1,-1}(0,\al)=-2\pi\nu \cot{\textstyle\frac{\pi}{2\nu}}(\nu\alpha+1) 
D_{1,b-2}(0,\al),
\quad\ \ \ b\ne 1\,,\label{basic-id1}\\
&D_{a,1}(1,\al)D_{1,-1}(0,\al)=-2\pi\nu \cot{\textstyle\frac{\pi}{2\nu}}(\nu\alpha+1) 
D_{a+2,-1}(0,\al),
\quad\quad a\ne -1\,,\label{basic-id2}\\
&D_{-1,1}(1,\al)D_{1,-1}(0,\al)=-\( 2\pi\nu\cot{\textstyle\frac{\pi}{2\nu}}(\nu\alpha+1)  \)^2\,.\label{basic-id3}
\end{align}
The equation \eqref{basic-id} for $a=1$, $b=-1$ is nothing but \eqref{1st-id} 
for $m=2$. 
By induction, \eqref{1st-id} for 
general $m$ reduces to \eqref{basic-id} and the elementary identity of determinants
\begin{align}
A_{1,1}^{m-2}\det (A_{i,j})_{1\le i,j\le m}
= \det\left(
\Bigl|\begin{matrix}
A_{1,1} & A_{1,j}\\
A_{i,1} & A_{i,j}\\
\end{matrix}\Bigr|
\right)_{2\le i,j\le m}\,.
\label{elementary}
\end{align}
Consider now \eqref{2nd-id} for $m=1$. 
If $1\notin I^-\cup \bar{I}^+$, then 
\eqref{2nd-id} is a consequence of 
\eqref{basic-id} and \eqref{elementary}. Else contractions
occur and one has to use together with \eqref{basic-id}
the identities \eqref{basic-id1}-\eqref{basic-id3}. This
is rather straightforward.
The case of general $m$ then follows by induction. 

Now we would like to discuss an analogue of the main formula
\eqref{themain} for the case of negative $m$. Consider first
of all the formula \eqref{basic-id3}. Supposing that
$\al,\al -2\frac{1-\nu}\nu\in (0,2)$ we can rewrite it as
\begin{align}
D_{-1,1}(0,\al)D_{1,-1}(-1,\al)
=-\(2\pi\nu\cot{\textstyle\frac{\pi}{2}}(\alpha-
{\textstyle\frac{1}{\nu}})\)^2\,.\nn
\end{align}
This identity implies
$$
\frac{\langle \Phi_{\alpha-2\frac{1-\nu}{\nu}}(0)\rangle^\mathrm{sG}_R}
{\langle \Phi_{\alpha}(0)\rangle^\mathrm{sG}_R}
=-C_{-1}(\al)\mub ^{ -2\al+2\frac {1-\nu} \nu  }\frac i {2\pi\nu^2}D_{-1,1}(0,\al)\,.
$$
where we define $C_{-m}(\al)$ by the equation
\begin{align}
C_{-m}(\al)C_m(\al -2m{\textstyle \frac{1-\nu}{\nu}})=\nu^{2m}\prod\limits _
{j=1}^m \tan^2{\textstyle\frac{\pi}{2}}(\alpha-
{\textstyle\frac{j}{\nu}})\,.
\label{Cnegative}
\end{align}

Generally, it is not hard to derive from \eqref{id-theta} another identity:
\begin{align}
&\Theta^{\mathrm sG}_R(l,j|\alpha-2{\textstyle \frac{1-\nu}{\nu}})
-\Theta^{\mathrm sG}_R(l-
{\textstyle\frac{i}{\nu}},j+
{\textstyle\frac{i}{\nu}}|\alpha)\label{id-theta1}\\
&\qquad=-\frac{\Theta^{\mathrm sG}_R(l-\frac{i}{\nu},\frac{i}{2\nu}|\alpha)
\Theta^{\mathrm sG}_R(-\frac{i}{2\nu},j+\frac{i}{\nu}|\alpha)}
{2\pi\nu\cot\frac{\pi}{2}(\alpha-\frac{1}{\nu})+\Theta^{\mathrm sG}_R(-\frac{i}{2\nu},\frac{i}{2\nu}|\alpha)
}\,,
\nn
\end{align}
which is understood as analytical continuation from the region 
$\al,\al -2\frac{1-\nu}\nu\in (0,2)$.
From \eqref{id-theta1} we obtain analogues of
\eqref{basic-id}-\eqref{basic-id3}, and further of 
\eqref{1st-id}, \eqref{2nd-id}:  
\begin{align}
&\det\left(D_{a,b}(0,\al)\right)_{a\in -\Io, b\in \Io
}=\prod_{j=0}^{m-1}D_{-1,1}(-j,\al),
\nn\\
&\det\left(D_{a,b}(0,\al)\right)_{a\in J^+, b\in J^-
%I_0(m)
}=(-1)^{\#}\prod\limits _{a\in (2m-I^+)_>\sqcup(2m-\bar{I}^-)_>}
2\pi\nu\cot{\textstyle\frac{\pi}{2\nu}}(\nu\alpha-a)
\nn\\&\qquad\quad
\times\det\left(D_{a,b}(m,\al)\right)_{a\in I^+\cup ( -\bar{I}^+)
\atop b\in I^- \cup (-\bar{I}^-)}\ \prod_{j=0}^{m-1}D_{-1,1}(-j,\al)\,,
\nn
\end{align}
where
\begin{align}
&J^+=(I^+-2m)_>\cup (-\bar{I}^+-2m)\cup (\Io \backslash (2m-
\bar{I}^-)_>) \,,\nn\\
&J^-= (I^-+2m)\cup (-\bar{I}^- + 2m)_<\cup (-\Io\backslash ({I}^+-2m)_<)\,,\nn
\end{align}
where $m>0$,  $\#(I^-)=\#(I^+)+m$, $\#(\bar{I}^+)=\#(\bar{I}^-)+m$.
So there is complete symmetry between positive and negative $m$. Then without going into the constructive definition of $\Phi _{\al}^{(m)}(0)$ we can just accept the validity of 
\eqref{eqtoprove} for all $m\in\mathbb{Z}$.
%%%%%%%%%%%%%%%%%%

\section{Comparison with known results}

\subsection{Lukyanov-Zamolodchikov formula}\label{LZformula}

Consider the case $R=\infty$. Notice that $\Theta^{\mathrm{sG}}_{\infty}(l,j|\al)=0$, so, in the 
formula \eqref{themain} the determinant contains a diagonal matrix. This means,
in particular, that $\langle \mathbf{l}_{-2}\Phi _\al(0)\rangle ^{\mathrm{sG}}_\infty=0$, and
the first descendant with non-trivial one-point function is $\mathbf{l}_{-2}\bar{\mathbf{l}}_{-2}\Phi _\al(0)$
as it has been expected.

Let us consider the simplest ratio of one-point functions of primary fields. 
The formula \eqref{themain} together with \eqref{musol} gives:
\begin{align}
&\frac{\langle\betab ^*_1
%\bar{\betab} ^*_{\mathrm{screen}, 1}
\bar{\gammab} ^*_{1}
%\gammab ^*_{\mathrm{screen},1}
\ \Phi ^{(1)}_{\al}(0)\rangle^{\mathrm{sG}}_\infty}
{\langle\Phi _{\al}(0)\rangle^{\mathrm{sG}}_\infty}
%\nn\\&
=-\frac i \nu 
\left[ M\frac {\sqrt{\pi} \ \Gamma (\frac 1 {2\nu})}
{2\Gamma (\frac {1-\nu}{2\nu})  }  \right]^{2(\nu\al +1-\nu)}\Gamma (\nu)^
{-2(\al+\frac {1-\nu}{\nu})}
\cot \frac {\pi}{2\nu}(\al \nu +1)%\tan\frac {\pi\nu} 2(\al -2)
\,.\nn
\end{align}
So, comparing with the formulae \eqref{C1} and \eqref{TWOPRI} we obtain
$$\frac{\langle \Phi _{\al +2\frac {1-\nu}\nu}(0)\rangle^{\mathrm{sG}}_\infty}
{\langle \Phi _{\al}(0)\rangle^{\mathrm{sG}}_\infty}=
\left[ M\frac {\sqrt{\pi} \ \Gamma (\frac 1 {2\nu})}
{2\Gamma (\frac {1-\nu}{2\nu})  }  \right]^{2(\nu\al +1-\nu)}
H(\al/2+(1-\nu)/2\nu)\,,$$
where
$$H(x)=\frac {\Gamma (-2\nu x)}
{\Gamma (2\nu x)}\cdot\frac{  \Gamma (x)  }{\Gamma (x+1/2 )}
\cdot\frac{\Gamma (-x+1/2 )}{  \Gamma (-x)  }\,.$$
This formula is in perfect agreement with the Lukyanov-Zamolodchikov formula
\cite{LukZam} which reads in our
notations as
\begin{align}
\langle \Phi _{\al}(0)\rangle^{\mathrm{sG}}_\infty&=
\left[ M\frac {\sqrt{\pi} \ \Gamma (\frac 1 {2\nu})}
{2\Gamma (\frac {1-\nu}{2\nu})  }  \right]^{ \frac {\nu ^2\al ^2}{2(1-\nu)  }}
\nn\\&\times\exp\Bigl(  \int\limits _0^{\infty}\Bigl(
\frac {\sinh ^2(\nu\al t)}{2\sinh (1-\nu)t \sinh t \cosh \nu t}-
\frac {\nu ^2\al^2}{2(1-\nu)} e^{-2t}\Bigr)\frac {dt}t \Bigr)\,.\nn
\end{align}

\subsection{Fateev-Fradkin-Lukyanov-Zamolodchikov-Zamolodchikov formula.}

Let us consider the first non-trivial
descendant for
$R=\infty$, which is $\mathbf{l}_{-2}\bar{\mathbf{l}}_{-2}\Phi _\al(0)$.
According to \eqref{descen}, 
\eqref{themain} and \eqref{musol}
we have:
\begin{align}
&\(D_1(\al)D_1(2-\al)\)^2
\frac {\langle \mathbf{l}_{-2}\bar{\mathbf{l}}_{-2}\Phi _{\al}(0)\rangle^{\mathrm{sG}}_\infty}
{\langle \Phi _{\al}(0)\rangle^{\mathrm{sG}}_\infty}\nn\\&=
\left[ M\frac {\sqrt{\pi} \ \Gamma (\frac 1 {2\nu})}
{2\Gamma (\frac {1-\nu}{2\nu})  }  \Gamma (\nu)^{-\frac 1 \nu}\right]^4\frac {1}{\nu^2}\cot\frac{\pi} {2\nu}(\nu\al-1)
\cot\frac{\pi} {2\nu}(\nu\al+1)\,.
\end{align}
Rewriting $D_1(\al)D_1(2-\al)$ by using \eqref{DE} we obtain
\begin{align}
\frac {\langle \mathbf{l}_{-2}\bar{\mathbf{l}}_{-2}\Phi _{\al}(0)\rangle^{\mathrm{sG}}_\infty}
{\langle \Phi _{\al}(0)\rangle^{\mathrm{sG}}_\infty}&=
-\left[ M\frac {\sqrt{\pi} \ \Gamma (\frac 1 {2\nu})}
{2\sqrt{1-\nu}\ \Gamma (\frac {1-\nu}{2\nu})  } \right]^4\nn\\&\times
\frac{ \Gamma (-\frac 1 2 +\frac {\al}2   +\frac 1 {2\nu} ) \Gamma (\frac 1 2 -\frac {\al}2   +\frac 1 {2\nu} )\Gamma (1- \frac {\al}2   -\frac 1 {2\nu} )\Gamma ( \frac {\al}2   -\frac 1 {2\nu} )}
{ \Gamma (\frac 3 2 -\frac {\al}2   -\frac 1 {2\nu} ) \Gamma (\frac 1 2 +\frac {\al}2   -\frac 1 {2\nu} )
\Gamma (\frac {\al}2   +\frac 1 {2\nu} ) \Gamma (1- \frac {\al}2   +\frac 1 {2\nu} )}\,.\nn
\end{align}
We want to compare this with the formula (1.8) of \cite{FFLZZ}.
The parameters are identified as follows:
$\xi =\frac {1-\nu}\nu$, $\eta =\al-1$. Making this change of variables  we find a perfect agreement.

\subsection{Zamolodchikov formula} \label{al=0}

In \cite{zamTT} A. Zamolodchikov proves that for any two-dimensional
Eucledian QFT on a cylinder the following formula holds:
\begin{align}
%\langle T_{z,z}T_{\bar{z},\bar{z}}\rangle\ 
%= \ \langle T_{z,z}  \rangle\langle T_{\bar{z},\bar{z}}  \rangle\ -\ \langle T_{z,\bar{z}}  \rangle^2
\langle T \bar T \rangle\ 
= \ \langle T \rangle\langle \bar T \rangle\ -\ \langle \Theta \rangle^2
\,,
\label{zam}
\end{align}
where $T=-2\pi T_{z,z}$,  $\bar T=-2\pi T_{\bar z,\bar z}$ and  $\Theta=2\pi T_{z,\bar z}$
are the components of the normalized energy-momentum tensor.

Let us check that the formulae \eqref{themain} agree with \eqref{zam}. 
We consider the
sG theory
with modified energy-momentum tensor. Obviously, considering \eqref{zam} we have to
set $\al =0$, so, \eqref{zam} reads
\begin{align}
\langle \mathbf{l}_{-2}\bar{\mathbf{l}} _{-2} \cdot 1\rangle ^\mathrm{sG}_R=
\langle \mathbf{l}_{-2}\cdot 1\rangle ^\mathrm{sG}_R
\ \langle \bar{\mathbf{l}} _{-2} \cdot 1\rangle ^\mathrm{sG}_R-\Bigl(2\pi\nu\frac{\mub^2}{\sin\pi\nu}
\Bigr) ^2
\(\langle \Phi_{2\frac {1-\nu}\nu}\rangle ^\mathrm{sG}_R\)^2\,,\label{zam1}
\end{align}
where the multiplier $2\pi\nu$ in the last term takes into account the CFT normalisation 
of the energy-momentum tensor and the scaling dimension of
$\mub$.

The case $\al=0$ is special because the singularity of $\hat{R}(k,0)$ at $k=0$ 
cancels.
From \eqref{R=R} we obtain
\begin{align}
\Theta ^{\mathrm{sG}}_R(l,j|0)=\Theta ^{\mathrm{sG}}_R
(-l,-j|0)\label{sym0}
\end{align}

The formula \eqref{zam1} follows immediately from the particular cases of \eqref{themain} :
\begin{align}
&\langle \mathbf{l}_{-2}\bar{\mathbf{l}} _{-2} \cdot 1\rangle ^\mathrm{sG}_R%\nn\\&
=\frac {M^4}{64 \nu ^2}
\left|\ \begin{matrix}    \Theta ^{\mathrm{sG}}_R
(\frac i{2\nu},\frac i{2\nu}|0)
& \Theta ^{\mathrm{sG}}_R(\frac i{2\nu},-\frac i{2\nu}|0)-2\pi\nu \cot\frac {\pi} {2\nu}
 \\ & \\
\Theta ^{\mathrm{sG}}_R
(-\frac i{2\nu},\frac i{2\nu}|0)-2\pi\nu \cot\frac {\pi} {2\nu} &\Theta ^{\mathrm{sG}}_R
(-\frac i{2\nu},-\frac i{2\nu}|0)
\end{matrix}
\ \right|\label{zamolo}\,,\\
\nn \\
&\langle \mathbf{l}_{-2}\cdot 1\rangle ^\mathrm{sG}_R=\frac{M^2}{8\nu}\Theta ^{\mathrm{sG}}_R
({\textstyle\frac i{2\nu}},{\textstyle\frac i{2\nu}}|0)\,,\quad
\langle \bar{\mathbf{l}}_{-2}\cdot 1\rangle ^\mathrm{sG}_R=
\frac{M^2}{8\nu}\Theta ^{\mathrm{sG}}_R
(-{\textstyle\frac i{2\nu}},-{\textstyle\frac i{2\nu}}|0)\,,\nn\\
&2\pi\nu\frac{\mub^2}{\sin\pi\nu}
\langle \Phi_{2\frac {1-\nu}\nu}\rangle ^\mathrm{sG}_R=\frac {M^2}{8\nu}
\(2\pi\nu \cot{\textstyle \frac {\pi} {2\nu}} -\Theta ^{\mathrm{sG}}_R
({\textstyle\frac i{2\nu}},-{\textstyle\frac i{2\nu}}|0)\)\,.\nn
\end{align}
From \eqref{sym0} we have $\Theta ^{\mathrm{sG}}_R
(\frac i{2\nu},-\frac i{2\nu}|0)=\Theta ^{\mathrm{sG}}_R
(-\frac i{2\nu},\frac i{2\nu}|0)$.

On the other hand there is
a simple independent way to compute the expectation values
of the components of the energy-momentum tensor. Let $E(R)$ 
be the energy of the ground state in Matsubara direction, corresponding momentum $P(R)$
equals zero. Then
\begin{align}
&
\langle T \rangle=
\langle \mathbf{l}_{-2}\cdot 1\rangle ^\mathrm{sG}_R
=\langle \bar T\rangle
=\langle \bar{\mathbf{l}}_{-2}\cdot 1\rangle ^\mathrm{sG}_R=
\frac 1 4\(\frac 1 R -\frac {d}{dR}\)E(R)\,,\nn\\
&\langle \Theta \rangle
=-2\pi\nu\frac{\mub^2}{\sin\pi\nu}
\langle \Phi_{2\frac {1-\nu}\nu}\rangle ^\mathrm{sG}_R
=\frac 1 4\(\frac 1 R +\frac {d}{dR}\)E(R)
\nn\,.
\end{align}

Let us see that they agree with the formulae \eqref{zamolo}. The check is based
on two equations:
\begin{align}
&\frac 1 i \frac{\partial}{\partial R}\log \mathfrak{a}(\z)=2\pi M (sh-R_\mathrm{dress}\ast  sh) (\z),\nn\\
&\frac 1 i \nu\z\frac{\partial}{\partial \z}\log \mathfrak{a}(\z)=2\pi M R (ch-R_\mathrm{dress}\ast  ch) (\z),\nn
\end{align}
where $sh (\xi)=\half (\xi ^{\frac 1\nu}-\xi ^{-\frac 1\nu})$, 
$ch (\xi)=\half (\xi ^{\frac 1\nu}+\xi ^{-\frac 1\nu})$.
Then using 
the formulae \cite{DDV}
\begin{align}
&4E(R)=
-2\pi M^2R\cot \frac {\pi} {2\nu} +
\frac{M}{2\pi\nu}\ 2\mathrm{Im}\int\limits _{0}^{\infty }(\xi ^{\frac 1 \nu}- \xi ^{-\frac 1 \nu}) \log
(1+\mathfrak{a}(\xi e^{+i 0}))\frac {d\xi ^2}{\xi ^2}
\,,
\end{align}
one immediately proves that
\begin{align}
 \(\frac 1 R -\frac {d}{dR}\)E(R)&=\frac{M^2}{2\nu}\(
e_+\ast R_\mathrm{dress}\ast e_+-e_+\ast e_+\)\nn\\&=\frac{M^2}{2\nu}\(
e_-\ast R_\mathrm{dress}\ast e_--e_-\ast e_-\)\,,\nn\\
\(\frac 1 R +\frac {d}{dR}\)E(R)&=\frac{M^2}{2\nu}\( 
e_+\ast R_\mathrm{dress}\ast e_--e_+\ast e_-\)-\pi M^2\cot \frac {\pi}{2\nu}\,,\nn
\end{align}
where $e_{\pm}(\xi)=\xi ^{\pm\frac 1 \nu}$. These equations are 
obviously equivalent to the last three equations of \eqref{zamolo}. 

\section{Conclusions}

The reader may notice certain discrepancy between simple final results of this paper 
and an indirect way of obtaining them in many cases.
One may find the definition of the lattice
regularisation of the temperature expectation values given in Section \ref{lattice} 
to be especially hard to understand. This situation reminds 
the history of writing the exact formulae for the form factors in 
the 
sG model. 
They were given in \cite{sm1986}, but the derivation was based
on the quantum Gelfand-Levitan-Marchenko equations. These
equations were derived using the lattice regularisation, 
and both their derivation and application are
rather a matter of art than that of science. However, later in \cite{KS} it was explained that the same formulae can be derived starting from the bootstrap approach which follows from
the first principles of QFT. We hope that the situation is similar
in the present case. Let us summarise the situation and 
%to analyse it.
analyse it.

We start with CFT, and we want to describe the quotient spaces
$\mathcal{V}^\mathrm{quo}_{\al+2m\frac{1-\nu}\nu}$. We claim that this can be done using the operators $\betab^*(\la)$,
$\gammab^*(\la)$,  $\gammab^*_{\mathrm{screen}}(\la)$ 
%create 
and creating
these spaces 
starting from $\phi _{\al}(0)$. The 
three-point functions with primary fields $\phi _{1-\kappa}(-\infty)$,
$\phi _{1+\kappa}(\infty)$ %is 
are
described by functions 
$\omega^\mathrm{sc} (\la,\mu|\kappa,\kappa, \al)$ and 
%$\omega (\la/\mu|\al)$. 
$\omega_0(\la/\mu,\al)$. Moreover, the very
meaning of our construction implies that similar formulae hold
%for 
in the case when the primary fields $\phi _{1-\kappa}(-\infty)$,
$\phi _{1+\kappa}(\infty)$ are replaced by any eigenvectors of 
the
local integrals of motion. It must be possible to make 
%this definitions all that 
these definitions
directly in CFT without any reference to the lattice. This should
be just a part of understanding the integrable structure of CFT.

The next statement concerns the integrable deformation of CFT.
We claim that the local fields in 
the
sG model are created by
the operators 
\begin{align*}
&\betab ^{+*}(\z)
=\betab^*(\mub\z)+\bar{\betab}^*_\mathrm{screen}
(\z/\mub)\,,
\\&
\gammab^{+*}(\z)
=\gammab^*(\mub\z)+\bar{\gammab}^*_\mathrm{screen}
(\z/\mub)\,,\\
&\betab ^{-*}(\z) =\bar{\betab}^*(\z/\mub)+{\betab}^*_\mathrm{screen}
(\mub \z)\,,\\
&\gammab^{-*}(\z) =\bar{\gammab}^*(\z/\mub)+\gammab^*_\mathrm{screen}
(\mub \z)\,,
\end{align*}
which provide 
a basis of local fields consistent with CFT (without
finite counterterms). 
The one-point functions
of these operators are computed using 
$\omega _R^\mathrm{sG}(\z,\xi|\al)$and 
$\omega_0(\z/\xi,\al)$.
It must be possible to 
explain this fact without reference to the lattice, by proper
understanding of an integrable deformation  
of CFT in its 
integrable formulation.

\appendix

\section{Three point function from CFT} \label{app1}

The three-point functions in CFT with $c<1$ were computed
by Dotsenko and Fateev \cite{DF}. 
Here it will be convenient for us to use the nice formula
obtained for Liouville theory by Zamolodchikov and 
Zamolodchikov \cite{ZZ3point}. 
We shall only need not the three-point function itself 
but rather its 
%Notice that in comparison with the fermionic formulas
%obtained in the previous sections we shall need
%not a particular  three-point functions, but rather 
ratio to the one with a shifted parameter. %two shifted ones. 
Teschner \cite{Tesc} proved the formula of \cite{ZZ3point}
computing exactly this kind of ratio. Another remark is %to be done
in order here. We apply the formulae obtained for Liuoville theory
to $c<1$ model. %In principle this 
This procedure involves 
some problems because of difficulties with understanding the
basic function $\Upsilon (x)$ for $b ^2<0$ (the notation is
%s are
given later). However, these problems do not concern
the ratios of three-point functions which we consider, and
which are expressed in terms of $\Gamma$-functions.
Finally, we normalise the primary fields as Liouville
exponential fields, i.e. they have non-trivial constant
in the two-point function.
For the last point, we refer the reader 
to Appendix C of \cite{AlZam}.
%see the Appendix C of \cite{AlZam}
%for relevant discussion.
We change the normalisation of the cosmological constant
of Liouville model comparing to \cite{ZZ3point}:
\begin{align}
\mathcal{A}^{
\mathrm{Lv}}=
\int \left\{\frac 1 {4  \pi} %\frac 1 {16  \pi} 
\partial _z\vphi (z,\bar{z})\partial_{\bar{z}}\vphi (z,\bar{z})+
\frac{\mub ^2}{\sin\pi b ^2} e^{b\vphis(z,\bar{z})}   \right\}
\frac{idz\wedge d\bar{z}}2\,,
%dzd\bar{z}\,.
\nn
\end{align}
Consider the three-point function of exponential fields.
%\textcolor{blue}{
In our cylindrical coordinates (recall that the map to
the Riemann sphere is given by $z\to e^{-z}$) it reads 
\begin{align}
\langle e^{a_1\vphis(-\infty)}e^{a_2\vphis(z,\bar{z})}
e^{a_3\vphis(\infty)}\rangle=
e^{(\Delta _{a_3}-\Delta _{a_1})(z+\bar{z})}
%|x_{1,2}|^{2\gamma _3}|x_{2,3}|^{2\gamma _1}
%|x_{3,1}|^{2\gamma _2}
C(a_1,a_2,a_3)\,,\label{THREEPOINT} 
\end{align}%}
where %$\gamma_1=\Delta _{a_1}-\Delta _{a_2}-\Delta _{a_3}$,
%{\it etc.}, and we have set
$$\Delta _a =a(Q-a),\quad Q=b+1/b.$$

We have from \cite{ZZ3point}
\begin{align}
C(a, Q/2-k,Q/2+k)&=
( \mub ^2\Gamma (1+b^2)^2\ b^{-2-2b^2})^{-a/b}%R^{-2a(Q-a)-Q^2+4k^2}
\nn\\&
\times \frac{\Upsilon_0
\Upsilon(2a)\Upsilon(Q-2k)\Upsilon (Q+2k)}{
\Upsilon ^2(a)\Upsilon (a-2k)\Upsilon(a+2k)}\nn\,.
\end{align}  
Following  \cite{ZZ3point} we use the function
$\Upsilon(x)$ defined by the equation
$$
\frac {\Upsilon(x+b)}{\Upsilon(x)}=\gamma(bx)b^{1-2bx},
\quad
\frac{\Upsilon(x+1/b)}{\Upsilon(x)}=\gamma(x/b)b^{-1+2x/b},
%\quad \gamma (x)=\frac{\Gamma(x)}{\Gamma(1-x)}\,,
$$
where $\gamma (x)=\frac{\Gamma(x)}{\Gamma(1-x)}$, 
as well as the constant $\Upsilon_0=(d\Upsilon/dx)(0)$.

An important point is that the exponent of the
%which we discussed is that the exponent of the
primary field must be of the sign opposite to that of the exponent in
the Lagrangian.   So, we consider the shift:
\begin{align}
\frac {C(a-b, Q/2-k,Q/2+k)}{C(a, Q/2-k,Q/2+k)}=
 \mub ^2\Gamma (1+b^2)^2
\frac {\gamma^2(ab-b^2)}{\gamma(2ab -2b^2)\gamma(2ab -b^2)}\label{xx}
\\
\times\gamma (ab-b^2-2kb)\gamma (ab-b^2+2kb)\,.
\nn
\end{align}
Now we go to our case of ``complex Liouville theory" by the substitution
%"complex Liouville theory" by substitution:
$$ 
b^2=\nu -1,\quad  2ab =\nu\al,\quad 2kb =\nu\kappa\,,$$
and obtain \eqref{3pt} in Section \ref{two}.
%Then the right hand side of \eqref{xx} turns into
%%The result is:
%\begin{align}
%&\frac {C(a-b, Q/2-k,Q/2+k)}{C(a, Q/2-k,Q/2+k)}
%%\label{xx}\\
%&%\to \   
%\mub ^2 \Gamma (\nu)^2\cdot Y(\al/2+(1-\nu)/2\nu)\ W(\al,\kappa)\overline{W}(\al,\kappa)\,,
%\label{xx1}
%\end{align}
%where
%\begin{align*}
%&Y(x)=2\nu x\cdot\frac {\Gamma ^2(\nu x +1/2-\nu/2)\Gamma (\nu-2\nu x)}
%{\Gamma ^2(1/2+\nu/2-\nu x)\Gamma (2\nu x +1-\nu)}
%\cdot\frac {\Gamma (-2\nu x)}{\Gamma (2\nu x)}\,,\\
%&W(\al,\kappa)=\frac{\Gamma(\al\nu/2-\nu+1+\kappa\nu)}{\Gamma(-\al\nu/2+\nu+\kappa\nu)}\,,\nn\\
%&\overline{W}(\al,\kappa)=W(\al,-\kappa)\,.\nn
%\end{align*}

\section{Asymptotic expansion of $\Theta$}\label{app2}

The first few terms of the asymptotic expansion 
of the function $\Theta(l,m|\kappa,\alpha)$ as $\kappa\to\infty$
are given as follows.  
We set $p=\kappa/\sqrt{8(1-\nu)}$,
\begin{align*}
&\Theta(il,im|\kappa,\alpha)
=\Theta^{\rm even}(il,im|\kappa,\alpha)+\Theta^{\rm odd}(il,im|\kappa,\alpha)d_\alpha\,,
\\
&\Theta^{*}(il,im|\kappa,\alpha)=\sum_{n=0}^\infty
\Theta^{*}_{2n}(il,im|\kappa,\alpha)p^{-2n}\,,
\end{align*}
where
\begin{align*}
d_\alpha&=\frac{\nu(\nu-2)}{\nu-1}(\alpha-1)\,.
\end{align*}
%%%%%%%%%%%%%%%%%%%%%%%%%%%%%%%%%%%%%%%%%%%%%%%%%%%%%%%%%%%%%%%%%%%%%%%%%%%%%%%%%%%%%%%%%%%%%%%%%%%
\begin{align*}
&\Theta^{\rm even}_0(il,im|\kappa,\alpha)=-\frac1{l+m}\,,
\\
&\Theta^{\rm even}_2(il,im|\kappa,\alpha)
=
\frac{{\Delta_\alpha}}{24 \nu }+\frac{2 [l+m]  \nu -1}{48 \nu }\,,
\\
&\Theta^{\rm even}_4(il,im|\kappa,\alpha)
=\frac{\Delta_\alpha^2}{1152\nu ^3}{\Bigl(-[l+m] \nu +1\Bigr)}
\\
&-\frac{\Delta_\alpha}{5760 (\nu -1) \nu^3}
\Bigl(\bigl[2\left(7 l^2+8 m l+7 m^2\right)+4 (l+m)\bigr]\nu ^3
\\
&\quad+\bigl[2 \left(7 l^2+8m l+7 m^2\right)+27 (l+m)+4
\bigr] \nu ^2-\bigl[27(l+m)+16\bigr] \nu +16\Bigr)
\\
&-\frac1{23040 (\nu -1) \nu^3}\Bigl(\bigl[28 (l+m)^3+8\left(l^2+3 m l+m^2\right)\bigr] \nu^4
\\
%\end{align*}
%\vskip -.7cm
%\begin{align*}
&\quad+\bigl[28 (l+m)^3+4\left(20 l^2+39 m l+20m^2\right)+16 (l+m)\bigr] \nu^3\\
&\quad-\bigl[4 \(20 l^2+39 m l+20m^2\right)+69(l+m)+6\bigr]
 \nu ^2+\bigl[69(l+m)+18\bigr] \nu -18\Bigr)\,,\\
&\Theta^{\rm even}_6(il,im|\kappa,\alpha)
=\frac{\Delta_\alpha^3}{82944 \nu^5}{\Bigl([(l+m)^2] \nu ^2-3 [l+m] \nu+2\Bigr) }
\\
&+\frac{\Delta_\alpha^2}{5806080 (\nu-1) \nu^5}
\Bigl(\bigl[126 (l+m)\left(3 l^2+2 m l+3 m^2\right)+8\left(13 l^2+50 m l+13m^2\right)\bigr] \nu ^4
\\
&\quad-\bigl[126\left(3 l^2+2 m l+3 m^2\right)(l+m)+9 \left(143 l^2+270 m l+143 m^2\right)+408 (l+m)\bigr]
\nu^3\\
&\quad+\bigl[9 \left(143 l^2+270 m l+143 m^2\right)+1773 (l+m)+256\bigr]\nu ^2\\&\quad
-\bigl[1773 (l+m)+774\bigr] \nu+774\Bigr)
\\
&+\frac{\Delta_\alpha}{3870720 (\nu -1)^2 \nu^5}
\Bigl(\bigl[4\left(93 l^4+208 m l^3+310 m^2 l^2+208m^3 l+93 m^4\right)\\
&\quad+328(l+m) \left(l^2+m l+m^2\right)+32\left(3 l^2+5 m l+3 m^2\right)\bigr] \nu^6
\\
&\quad-\bigl[8 \left(93 l^4+208 m l^3+310 m^2 l^2+208 m^3 l+93m^4\right)
\\
&\quad+4 \left(610 l^2+561 m l+610m^2+68\right) (l+m)+8 \left(183 l^2+274 m l+183 m^2\right)\bigr] \nu ^5\\
&\quad+\bigl[4 \left(93 l^4+208 m l^3+310 m^2 l^2+208 m^3 l+93m^4\right)\\
&\quad
+8(l+m) \left(528 l^2+479 m l+528m^2\right)\\
&\quad+5831 l^2+8422 m l+5831 m^2+2138 (l+m)+184\bigr] \nu^4
\\
&\quad-\bigl[4 \left(528 l^2+479 m l+528 m^2+1504\right)(l+m)+2 \left(4367 l^2+6230m l+4367 m^2\right)%+6024 (l+m)
\\
&\quad+1104\bigr] \nu^3+\bigl[4367 l^2+6230 m l+4367m^2+7772 (l+m)+2556\bigr]\nu^2\\
&\quad-\bigl[3886 (l+m)+2904\bigr] \nu +1452\Bigr)
%\\
\end{align*}
\begin{align*}
&
+\frac1{23224320 (\nu -1)^2 \nu ^5}
\Bigl(\bigl[744 (l+m)^5+16\left(41 l^4+205 m l^3+308 m^2 l^2+205m^3 l+41 m^4\right)\\
&\quad+192\left(l^2+4 m l+m^2\right) (l+m)\bigr] \nu^7\\
&\quad-\bigl[1488 (l+m)^5+4\left(1477 l^4+5990 m l^3+8986 m^2l^2+5990 m^3 l+1477 m^4\right)
\\
&\quad+24 \left(163l^2+413 m l+163 m^2\right) (l+m)+96\left(9 l^2+25 m l+9 m^2\right)\bigr]\nu ^6
\\
&\quad+\bigl[744 (l+m)^5+8 \left(1313l^4+5170 m l^3+7754 m^2 l^2+5170 m^3l+1313 m^4\right)
\\
&\quad+6 \left(2953l^2+6064 m l+2953 m^2+184\right)(l+m)+4 \left(2023 l^2+4538 m l+2023 m^2\right)
%+1104 (l+m)
\bigr] \nu^5\\
%\end{align*}
%\vskip -.7cm
%\begin{align*}
&\quad-\bigl[4\left(1313 l^4+5170 m l^3+7754 m^2l^2+5170 m^3 l+1313m^4\right)\\
&\quad+6 (l+m) 
\left(4602l^2+8824m l+4602 m^2+1125\right)\\&\quad
+24755 l^2+50326 m l+24755m^2+360\bigr] \nu ^4\\
&\quad+\bigl[6\left(2301 l^2+4412 m l+2301m^2+2460\right) (l+m)\\
&\quad+2\left(16663 l^2+32174 m l+16663m^2\right)
%+15840 (l+m)
+1800\bigr] \nu^3
\\
&\quad-\bigl[ 16663 l^2+32174 m l+16663m^2+18180 (l+m)+3600\bigr] \nu^2\\
&\quad+\bigl[ 9090 (l+m)+3600\bigr] \nu-1800\Bigr)\,,\\
%\end{align*}
%Coefficient of $d_\alpha p^{-6}$
%\begin{align*}
&\Theta^{\rm odd}_0(il,im|\kappa,\alpha)=0\,,
\\
&\Theta^{\rm odd}_2(il,im|\kappa,\alpha)=0\,,
\\
&\Theta^{\rm odd}_4(il,im|\kappa,\alpha)
=-\Delta_\alpha\frac{l-m}{2880 \nu ^2}\,, 
\\
&\Theta^{\rm odd}_6(il,im|\kappa,\alpha)
=\Delta_\alpha^2\frac{l-m}{69120 \nu^4}{([l+m] \nu -2)}
\\
&+\Delta_\alpha\frac{l-m}{2903040 (\nu -1) \nu^4}
\Bigl(\left(2\left[41 l^2+42 m l+41m^2\right)+48 (l+m)\right] \nu ^3
\\
&\quad -\left[2 \left(41 l^2+42 m l+41m^2\right)+285(l+m)+96\right]\nu ^2
+\bigl[285(l+m)+302\bigr] \nu -302\Bigr)
\,.
\end{align*}

%%%%%%%%%%%%%%%%%%%%%%%%%%%%%%%%%%%%%%%%%%%%%%%%%%%%%%%%%%%%%%%%%%%%%%%%%%%%%%%%%%%%%%%%%%%%%%%%%%%

Expectation values of the fermions $\betab^*_{2j-1}$, etc. are calculated by 
appropriate specialisation of $l,m$. 
We give below the results upto degree $\kappa^{-8}$
in the case 
\begin{align*}
\Omega_{2r-1,2s-1}(\kappa,\alpha)
&=-{\textstyle\frac{r+s-1}{\nu}}(\sqrt{2} p \nu)^{2r+2s-2}
\Theta\left({\textstyle\frac{i(2r-1)}{2\nu} ,\frac{i(2s-1)}{2\nu}}
\Bigl|
\kappa,\alpha\right)
%\left(\frac{\sqrt{2} \ f\kappa \nu}{R}\right)^{2r+2s-2}\,,
\\
&=\Omega^{\rm even}_{2r-1,2s-1}(\kappa,\alpha)+\Omega^{\rm odd}_{2r-1,2s-1}(\kappa,\alpha)
d_\alpha\,.
\end{align*}
%where
%\begin{align*}
%d_\alpha&=\frac{\nu(\nu-2)}{\nu-1}(\alpha-1)\,.
%\end{align*}
In the next formulae, 
$I_{2n-1}(\kappa)$'s denote the vacuum eigenvalues of the integrals of motion
which can be found for instance in \cite{BLZI}.
\begin{align*}
\Omega^{\rm even}_{1,1}(\kappa,\al)
&=
I_1(\kappa)-\frac{\Delta_\alpha}{12},
\\
\Omega^{\rm even}_{1,3}(\kappa,\al)
&=
I_3(\kappa)
-\frac{\Delta_\alpha}{6}I_1(\kappa)
+\frac{\Delta_\alpha^2}{144}
+\frac{c+5}{1080}\Delta_\alpha\,,
\\
\Omega^{\rm odd}_{1,3}(\kappa,\al)
&=- \frac{\Delta_\alpha}{360}\,,
\end{align*}
\begin{align*}
\Omega^{\rm even}_{1,5}(\kappa,\al)
&=
I_5(\kappa)
-\frac{\Delta_\alpha}{4}I_3(\kappa)
+\left(\frac{\Delta_\alpha^2}{48}
+\frac{c+11}{360}\Delta_\alpha
\right)I_1(\kappa)
\\
&\quad 
-\frac{\Delta_\alpha^3}{1728}
-\frac{13(c+35)}{90720}\Delta_\alpha^2
-\frac{2c^2+21c+70}{60480}\Delta_\alpha\,,\\
%\end{align*}
%\vskip -.7cm
%\begin{align*}
\Omega^{\rm even}_{3,3}(\kappa,\al)
&=
I_5(\kappa)
-\frac{\Delta_\alpha}{4} I_3(\kappa)
+\left(\frac{\Delta_\alpha^2}{48}+\frac{c+2}{360}
\Delta_\alpha+\frac{c+2}{1440}\right)I_1(\kappa)
\\
&\quad
-\frac{1}{1728}\Delta^3_\alpha
-\frac{5c-14}{18144}\Delta^2_\alpha
-\frac{10c^2+37c+70}{362880}\Delta_\alpha
-\frac{1/2c^2+c}{36288}\,,
\\
\Omega^{\rm odd}_{1,5}(\kappa,\al)
&=-\frac{\Delta^2_\alpha}{1440}
-\frac{c+7}{7560}\Delta_\alpha
-\frac{\Delta_\alpha}{120}I_1(\kappa)
\,.
\end{align*}

\bigskip

\noindent

{\it Acknowledgements.}\quad
Research of MJ is supported by the Grant-in-Aid for Scientific 
Research B-20340027. 
Research of TM is supported by the Grant-in-Aid for Scientific Research
B-22340031.
Research of FS is supported by   SFI
under Walton Profesorship scheme and by
RFBR-CNRS grant 09-02-93106
and by SFI
under Walton Profesorship scheme.
The authors would like to thank for the hospitality extended by Theory Group at DESY, Hamburg, 
where an important part of this work was done under the EU-grant MEXT-CT-2006-042695.
Special thanks are due to J. Teschner for many valuable discussions.

\bigskip
%%%%%%%%%%%%%%%%%%%%%%%%%%%%

\end{document}